\documentclass[lettersize,journal]{IEEEtran}
\IEEEoverridecommandlockouts
\usepackage{cite}
\usepackage{amsmath,amssymb,amsfonts}
\usepackage{graphicx}
\usepackage{textcomp}
\usepackage{xcolor}
\usepackage{dsfont}
\usepackage{comment}

\usepackage{hyperref}
\usepackage{amsmath}
\usepackage{amsfonts}
\usepackage{amssymb}
\usepackage{amsthm}
\newtheorem{theorem}{Theorem}

\newtheorem{lemma}{Lemma}

\newtheorem{assumption}{Assumption}

\newtheorem{problem}{Problem}
\newtheorem{remark}{Remark}
\usepackage{cite}
\usepackage{algorithm}
\usepackage{algpseudocode}
\usepackage{multirow}
\usepackage{multicol}
\usepackage{mathtools}

\DeclareMathOperator*{\argmax}{arg\,max}

\newcommand{\circled}[1]{\raisebox{.5pt}{\textcircled{\raisebox{-.9pt} {#1}}}}
\def\BibTeX{{\rm B\kern-.05em{\sc i\kern-.025em b}\kern-.08em
    T\kern-.1667em\lower.7ex\hbox{E}\kern-.125emX}}

\usepackage[utf8]{inputenc}
\usepackage{pgfplots}
\usepgfplotslibrary{groupplots,dateplot}
\usetikzlibrary{patterns,shapes.arrows}
\pgfplotsset{compat=newest}
\pgfplotsset{scaled y ticks=false}
\usepackage{caption}
\usepackage{subcaption}

\usepackage{soul}

\begin{document}
\title{A Community-Aware Framework for \\ Social Influence Maximization}

\author{Abhishek K. Umrawal, Christopher J. Quinn,~\IEEEmembership{Member,~IEEE}, and Vaneet Aggarwal,~\IEEEmembership{Senior Member,~IEEE} 
\thanks{A. K. Umrawal is with the School of Industrial Engineering, Purdue University, West Lafayette, IN, 47907, USA (email: \href{aumrawal@purdue.edu}{aumrawal@purdue.edu}), and Department of Computer Science and Electrical Engineering, University of Maryland, Baltimore County, Baltimore, MD, 21250, USA. C. J. Quinn is with the Department of Computer Science, Iowa State University, Ames, IA, 50011, USA (email: \href{cjquinn@iastate.edu}{cjquinn@iastate.edu}). V. Aggarwal is with the School of Industrial Engineering, and Elmore Family School of Electrical and Computer Engineering, Purdue University, West Lafayette, IN, 47907 USA (email: \href{vaneet@purdue.edu}{vaneet@purdue.edu}). He is also with Computer Science, KAUST, Thuwal, 23955, KSA.} \thanks{This material is based upon work supported in part by the National Science Foundation under Grants No. 1742847, 2149588, and 2149617.} \thanks{This paper has been accepted for publication in IEEE Transactions on Emerging Topics in Computational Intelligence (TETCI) in Dec 2022.}
\thanks{\copyright~2022 IEEE.  Personal use of this material is permitted.  Permission from IEEE must be obtained for all other uses, in any current or future media, including reprinting/republishing this material for advertising or promotional purposes, creating new collective works, for resale or redistribution to servers or lists, or reuse of any copyrighted component of this work in other works.}
}

\maketitle
\begin{abstract}
We consider the problem of \textit{Influence Maximization} (IM), the task of selecting $k$ seed nodes in a social network such that the expected number of  nodes influenced is maximized. We propose a community-aware divide-and-conquer framework that involves (i) learning the inherent community structure of the social network, (ii) generating candidate solutions by solving the influence maximization problem for each community, and (iii) selecting the final set of seed nodes using a novel progressive budgeting scheme. 

Our experiments on real-world social networks show that the proposed framework outperforms the standard methods in terms of run-time and the heuristic methods in terms of influence. We also study the effect of the community structure on the performance of the proposed framework. Our experiments show that the community structures with higher modularity lead the proposed framework to perform better in terms of run-time and influence.
\end{abstract}

\begin{IEEEkeywords}
Social networks, influence maximization, viral marketing, community detection, submodular maximization
\end{IEEEkeywords}
\section{Introduction} \label{section1}

\subsection{Motivation} \IEEEPARstart{T}{he} advent of social media has changed how traditional marketing strategies were used to be designed \cite{evans2010social}. Companies are now preferring to allocate a significant proportion of their marketing budget to drive sales through large social media platforms. There are several ways in which social media can be leveraged for promotional marketing. For instance, advertising on the most visited social platforms, making social media pages for branding and spreading the word about the product, etc. A more sophisticated approach for promotional marketing would be to use the dynamics of the social network to identify the right individuals to be incentivized to get the maximum influence in the entire network. 

In the context of social media marketing, Domingos and Richardson posed the \textit{Influence Maximization} (IM) problem \cite{domingos2001mining}: ``if we can try to convince a subset of individuals in a social network to adopt a new product or innovation, and the goal is to trigger a large cascade of further adoptions, which set of individuals should we target?'' Formally, it is the task of selecting $k$ seed nodes in a social network such that the expected number of influenced nodes in the network (under some influence propagation model), referred to as the \textit{influence}, is maximized. Kempe et al. \cite{kempe2003maximizing} showed that the problem of influence maximization is NP-Hard.  This problem has been widely studied in the literature and several  approaches for solving it have been proposed. Some approaches provide near-optimal solutions but are costly in terms of run time.  On the other hand, some approaches are faster but heuristics, i.e. do not have approximation guarantees. 

Motivated by the idea of addressing this trade-off between accuracy and run-time, we propose a community-aware divide-and-conquer framework to provide a time-efficient solution. The proposed framework outperforms the standard methods 
in terms of run-time and the heuristic methods in terms of influence.

\subsection{Literature Review} \label{subsec:lit_review}
Researchers have proposed different algorithms and heuristics for solving the Influence Maximization (IM) problem using several approaches. We now discuss several categories of the relevant approaches as follows.  We refer to methods that presume knowledge of the network and estimate influence using Monte Carlo simulations of the diffusion process as \textit{simulation-based methods}.  

\subsubsection{Simple heuristics} Degree centrality is perhaps the simplest way to quantify the influence of an individual in the network \cite{kempe2003maximizing}. Observing the fact that many of the most central nodes may be clustered, targeting all of them is not at all necessary, Chen et al. \cite{chen2009efficient} proposed the degree discount heuristic. These heuristics are simple and time-efficient. However, they do not have any provable guarantees.
    
\subsubsection{Simulation-based methods} 
Under the \textit{independent cascade}  \cite{goldenberg2001talk, goldenberg2001using} and \textit{linear threshold} \cite{granovetter1978threshold, schelling2006micromotives} models of diffusion (discussed in Section~\ref{sec:diffusion_models}), Kempe et al. \cite{kempe2003maximizing} showed that the problem of influence maximization is NP-Hard. They also proposed to use an efficient greedy algorithm \cite{domingos2001mining} which due to a result by Nemhauser et al. \cite{nemhauser1978analysis} gives an $\left(1-\frac{1}{e}\right)$-approximation  of the solution. The asymptotic run-time of this algorithm is O($nk$). Asymptotically, this greedy algorithm is efficient but empirically the costly Monte Carlo simulations cause an overhead. Leskovec et al. \cite{leskovec2007cost} proposed the CELF algorithm which improves upon the empirical run-time of the simple greedy algorithm by further exploiting the property of submodularity. Goyal et al. \cite{goyal2011celf++} proposed the CELF++ algorithm which further improved upon the empirical run-time of the CELF algorithm by even further exploiting the property of submodularity to avoid unnecessary re-computations of marginal gains incurred by CELF. Borgs et al. \cite{borgs2014maximizing} proposed a greedy algorithm using reverse influence sampling (RIS) -- an approach to efficiently estimate the influence of a seed set. CELF, CELF++, and \cite{borgs2014maximizing} have the same worst-case run time O($nk$) and approximation ratio $\left(1-\frac{1}{e}\right)$ as the one proposed by Kempe et al. \cite{kempe2003maximizing}. Lotf et al. \cite{lotf2022improved} proposed a genetic algorithm-based heuristic algorithm for dynamic (evolving over time) networks. This method involves Monte Carlo simulation and does not have any approximation guarantees. The framework proposed in this paper may also involve Monte Carlo simulations. But, the divide-and-conquer strategy allows us to significantly reduce the run-time.

\subsubsection{Community-based methods} As the proposed method utilizes the inherent community structure of the network, we discuss other community-based methods of influence maximization as follows. Chen et al. \cite{chen2012efficient} proposed two methods called CDH-KCut and CDH-SHRINK under heat diffusion model \cite{ma2008mining}. They further improved their methods and proposed another method called CIM \cite{chen2014cim}. Bozorgi et al. \cite{bozorgi2016incim} proposed a method called INCIM which works only for the linear threshold diffusion model. Moreover, the method involves overlapping community detection contrary to our work where the communities are non-overlapping. Bozorgi et al. \cite{bozorgi2017community} have also developed a method for competitive influence maximization \cite{10.1007/978-3-540-77105-0_31} under the competitive linear threshold model. Shang et al. \cite{shang2017cofim} have proposed a method called CoFIM under the independent cascade diffusion model and weighted cascade edge-weight model. Contrary to these methods, our method does not depend on the choice of the diffusion model. Huang et al. \cite{huang2019community} proposed a data-based method called CTIM which requires a potential action log and item-topic relevance.

\subsubsection{Data-based methods}
Provided some observational data involving real-world diffusion traces is available, the Monte Carlo simulations can be avoided by estimating the influence directly from the data. Goyal et al. \cite{goyal2011data}, instead of using a propagation model, proposed a data-based-method to introduce a model called the credit distribution model, which directly leverages the propagation traces from real-world data and learns the flow of influence in the network. Pen et al. \cite{pan2015credit} and Deng et al. \cite{deng2015credit} have studied variants of the credit distribution model under time constraints and node features respectively. The proposed method does not involve any observational data. 

\subsubsection{Online methods}
More recently, the focus has been on solving the problem of influence maximization in an online manner where the goal is to maximize the cumulative observed influence of the seed sets chosen at different times while receiving instantaneous feedback. Approaches differ  based on semi-bandit feedback \cite{lei2015online, wen2017online, vaswani2017diffusion, li2020online, perrault2020budgeted} and full-bandit feedback \cite{agarwal2022stochastic, nie2022explore}. The proposed method is not an online method. 

\subsection{Contribution} In Section~\ref{subsec:lit_review}, we discussed that the CELF++ \cite{goyal2011celf++} algorithm is faster compared to the simple greedy algorithm \cite{kempe2003maximizing, domingos2001mining}. But the costly aspect of performing a large number of diffusions in the entire network is still there. Motivated by the idea of solving the influence maximization problem in a time-efficient manner, we propose a community-aware divide-and-conquer framework that involves (i) learning the inherent community structure of the social network, (ii) generating candidate solutions by solving the influence maximization problem for each community, and (iii) selecting the final set of individuals to be incentivized from the candidate solutions using a novel progressive budgeting scheme. Our method may also use the Monte Carlo simulations but we are restricting them within each community as compared to the entire network which brings savings in terms of run-time as compared to the CELF++ algorithm.

Compared to the other community-based methods, the proposed framework is novel in the following ways. It is not limited to a specific diffusion and/or an edge-weight model. In Step 1, the set of candidate solutions is generated by all combinations of solutions from each community. In Step 2, the final seed selection is performed by solving an integer linear program (ILP) over candidate solutions subject to a budget constraint. We propose an efficient progressive budgeting scheme to efficiently solve the ILP in Step 3. We provide the proof of correctness of this scheme which leverages submodularity (defined in Section~\ref{section2}) of the influence.

We provide experiments on real-world social networks, showing that the proposed framework outperforms simulation-based methods in terms of run-time and heuristic methods in terms of influence. We study the effect of the community structure on the performance of the proposed framework. Our experiments show that the community structures with higher `modularity' (defined in Section~\ref{section2}) lead the proposed framework to perform better in terms of run-time and influence.

\subsection{Organization} The rest of the paper is organized as follows. In Section~\ref{section2}, we discuss the preliminaries and formulate the problem. In Section~\ref{section3}, we discuss our methodology. In Section~\ref{section4}, we discuss the experiments performed on real-world social networks. Section~\ref{section5} concludes the paper and provides future directions. 

\section{Preliminaries and Problem Formulation} \label{section2}
In this section, we discuss some preliminaries and formulate the problem of interest in this paper. Refer to Appendix~\ref*{sec:notations} for a table of important notations used throughout the paper.

\subsection{Submodularity}
Let $\Omega$ denote the ground set of $n$ elements and $2^\Omega$ denote the set of all subsets of $\Omega$. A set function $f: 2^{\Omega} \rightarrow \mathbb{R}$ is said to be \textit{submodular} if it satisfies a natural `diminishing returns' property: the marginal gain from adding an element $v$ to a set $S\subseteq \Omega$ is at least as high as the marginal gain from adding the same element $v$ to a superset $T\subseteq \Omega$ of $S$. Formally, for any sets $S,T \subseteq \Omega$ such that $S \subseteq T$, $f$ satisfies
\begin{align}
    f(S\cup\{v\})-f(S) \ge f(T\cup\{v\})-f(T).
\end{align} %
A set function $f : 2^{\Omega} \rightarrow \mathbb{R}$, is said to be \textit{monotone} (non-decreasing) if for any sets $S,T \subseteq \Omega$ such that $S \subseteq T$, $f$ satisfies \begin{align}
    f(S) \leq f(T).
\end{align}

\subsection{Diffusion models and social influence} \label{sec:diffusion_models}
There are several discrete-time stochastic models of diffusion over social networks. For the purpose of our research, we focus on the \textit{independent cascade} (IC) \cite{goldenberg2001talk, goldenberg2001using} and \textit{linear threshold} (LT) \cite{granovetter1978threshold, schelling2006micromotives} models of diffusion.

In the independent cascade model, given a graph $G=(V,E)$, the process starts at time $0$ with an initial set of active nodes $S$, called the \textit{seed set}. When a node $v \in S$ first becomes active at time $t$, it will be given a single chance to activate each currently inactive neighbor $w$, it succeeds with a probability $p_{v,w}$ (independent of the history thus far). If $w$ has multiple newly activated neighbors, their attempts are sequenced in an arbitrary order. If $v$ succeeds, then $w$ will become active at time $t+1$; but whether or not $v$ succeeds, it cannot make any further attempts to activate $w$ in subsequent rounds. The process runs until no further activation is possible. 

In the linear threshold model, given a graph $G=(V,E)$, a node $v$ is influenced by each neighbor $w$ according to a weight $p_{v,w}$ such that $\sum_{w \in \partial v} p_{v,w} \le 1$, where $\partial v$ represents the set of neighbors of $v$. Each node $v$ chooses a \textit{threshold} $\theta_v$ uniformly from the interval [0,1]; this represents the weighted fraction of $v$'s  neighbors that must become active in order for $v$ to become active. The process starts with a random choice of thresholds for the nodes, and an initial set of active nodes $S$, called the \textit{seed set}. In step $t$, all nodes that were active in step $t-1$ remain active, and we activate any node $v$ for which the total weight of its active neighbors is at least $\theta_v$. The process runs until no more activation is possible.

Note that both these processes of diffusion are \textit{progressive}, i.e. the nodes can switch from being inactive to active, but do not switch in the other direction. 

At any time $t$ in the cascade, each node $v \in V$ can be either active or inactive. We denote the process for each node $v\in V$'s state as $\{Y_t^{(v)}\}_{t=1}^T$
\begin{equation}
Y_t^{(v)} = \begin{cases} 
      1, & \text{if node $v$ is active at time $t$,} \\
      0, & \text{otherwise.} \\
   \end{cases}
\end{equation}

The \textit{influence} $\sigma(S)$ of a set $S$ is defined as the expected number of active nodes at the end of the cascade (denoted by time $T$), given that $S$ is the  set of initially active nodes,
\begin{align}
    \sigma(S) &= \mathbb{E}\left[\sum_{v \in V} Y_T^{(v)} \bigg |  \bigcap_{v\in V} \{Y_0^{(v)} = \mathds{1}(v\in S)\} 
    \right],
    \label{eq:equiv_def_influence}
\end{align} %
where $\mathds{1}(\cdot)$ denotes the indicator function.

Kempe et al. \cite{kempe2003maximizing} showed that under common models of diffusion such as independent cascade and linear threshold models, $\sigma(S)$ is a monotone non-decreasing submodular set function.

\subsection{Problem statement}
For a given integer budget $k$, we are interested in finding a $k$-node subset of the set of nodes $V$, which has the maximum influence over all possible $k$-node subsets of $V$. Formally, the problem of influence maximization (IM) is defined as
\begin{problem} \label{eq:main_problem}
\begin{align*} 
    & \argmax_{S \subseteq V} \ \sigma(S),\\
    \text{s.t.} & \qquad |S| \le k. \tag{budget constraint} 
\end{align*}
\end{problem}
\section{Methodology} \label{section3}
With the goal of solving the influence maximization problem (Problem~\ref{eq:main_problem}) in a time-efficient manner, we propose a community-aware divide-and-conquer framework.  The proposed framework reduces the search space for the seed sets by partitioning the given network using its inherent community structure.   The proposed framework involves (i) learning the inherent community structure of the social network, (ii) generating candidate solutions by solving the influence maximization problem for each community, and (iii) selecting the final set of individuals to be incentivized from the candidate solutions using a novel progressive budgeting scheme.

Algorithm~\ref{alg:community-im} outlines the framework proposed in this paper. It uses three sub-routines which are explained in the following subsections.

\begin{algorithm}
\caption{Community-IM}\label{alg:community-im}
\begin{algorithmic}[1]
\State \textbf{Input} {Graph $G$, budget $k$, \texttt{com-method}, \texttt{sol-method}}.
    \State $\{G_i\}_{i=1}^c\gets$  \text{Community-Detection}($G$, \texttt{com-method})
    \For {community $i = 1, \dots, c$} 
        \State $\mathcal{S}_i,\Sigma_i \gets$  \text{Generate-Candidates}($G_i, k$, \texttt{sol-method})
    \EndFor
    \State $S^{*} \gets$  \text{Progressive-Budgeting}($\{\mathcal{S}_i\}_{i=1}^c$,$\{\Sigma_i\}_{i=1}^c$,$k$)
    \State \textbf{return} $S^{*}$
\end{algorithmic}
\end{algorithm}

\subsection{Learning the  community structure of the network}  For the given social network $G=(V,E)$, we obtain a hard partition $\{V_1,\dots,V_c\}$ of the node set $V$ using some community detection method. By hard partitioning, we mean we mean the communities are non-overlapping, i.e. $V_i \cap V_j = \emptyset$ for all communities $i\neq j$ with $i,j\in\{1,\dots,c\}$ and $\bigcup_{i=1}^c V_i = V$. Define $G_i = (V_i,E_i)$ where $E_i$ is the set of edges from $E$ connecting pairs of nodes in $V_i$. We call $\{G_1,\dots,G_c\}$ a network-partition.

Most community detection methods select communities such that the nodes within a community are more `well-connected' than the nodes between communities. Methods  differ in how they explicitly or implicitly measure the connectedness of the nodes in a network. Common community detection methods are the Louvain  method \cite{blondel2008fast}, label propagation \cite{cordasco2010community}, and the Girvan-Newman algorithm \cite{girvan2002community}.

\subsubsection{Quality of a network-partition}
The quality of a network-partition can be measured using modularity score \cite{newman2018networks,clauset2004finding}.
The modularity score of a network-partition is defined as the fraction of the edges that fall within the given groups minus the expected fraction if edges were distributed at random. For a network-partition $\{G_1,\dots,G_c\}$, modularity \cite{clauset2004finding} is defined as
\begin{align*}
    Q &= \sum_{i=1}^{c}\left[\frac{L_i}{|E|} - \left(\frac{\delta_i}{|E|}\right)^2\right],
\end{align*}
where $L_i$ is the number of edges between the pairs of nodes in $G_i$ and $\delta_i$ is the sum of the degrees of nodes in $G_i$.

The modularity score is used as a measure of how well a community detection algorithm partitions a network. A higher value of modularity corresponds to a network-partition with higher connectedness within each community.

\subsubsection{Community detection methods} \label{sec:community-detection-methods}
We discuss some commonly used community detection methods (\texttt{com-method} in  Algorithm~\ref{alg:community-im}). The Louvain method \cite{blondel2008fast} first obtains small communities by optimizing modularity locally on all of the nodes.  Then each small community is treated as a single node and the previous step is repeated. Label propagation  \cite{cordasco2010community} starts with a (generally small) random subset of the nodes with  community labels. The algorithm then iteratively assigns labels to previously unlabeled nodes. The Girvan-Newman method \cite{girvan2002community} method uses a measure known as `betweenness.'  Define the betweenness of an edge \cite{girvan2002community} as the sum of the `weights' of the shortest paths between any pair of nodes that run along it. If there are $d$ different shortest paths between any two nodes then the weight of each path is set as $1/d$. The Girvan-Newman method \cite{girvan2002community} method involves the following steps.
\begin{enumerate}
    \item First, calculate the betweenness of all existing edges in the network.
    \item Next, remove the edge(s) with the highest betweenness. 
    \item Finally, recalculate the betweenness of all edges affected by the removal at the previous step. 
    \item Repeat the previous two steps until no edge remains.
\end{enumerate}

For the framework proposed in this paper, the only formal requirement for the community detection method (\texttt{com-method} in  Algorithm~\ref{alg:community-im}) is that it provides a hard partition. Based on our experiments (discussed in Section~\ref{section4}), we observe that the Louvain method \cite{blondel2008fast} works the best for our framework.

\subsection{Generating candidate solutions by solving the influence maximization problem for each community}
For each  community $G_i$, we find the best seed sets of sizes $1,\dots,k$ for that community using some standard influence maximization method. Let $S_{i,j}$ denote the best seed set of size $j$  for community $i$.  Let $\sigma_i(S_{i,j})$ denote  the corresponding expected influences of those seed sets within community $i$ ($i=1,\dots,c$).  

Solving the influence maximization problem separately for different communities  instead of the entire network improves the empirical run-time. The partitioning of the original network leads to fewer subset evaluations (oracle calls). Furthermore, those (fewer) evaluations are also faster to carry out. Refer to Appendix~\ref*{sec:complexitystep2} for details.

For the framework proposed in this paper, any standard influence maximization method can be used as \texttt{sol-method} in Algorithm~\ref{alg:community-im}. 
For our experiments (discussed in Section~\ref{section4}), we use the CELF++ method \cite{goyal2011celf++} to demonstrate our framework.

Later, to discuss guarantees of our method (on a surrogate optimization problem), we will assume that the \texttt{sol-method} used has the following property.

\begin{assumption} \label{asmptn:decreasing}
We assume that the marginal gains $\{\sigma_i(S_{i,j+1}) - \sigma_i(S_{i,j})\}_{j=1}^{k-1}$  within each community $i\in\{1,\dots,c\}$ are non-increasing.
\end{assumption}
\begin{remark} \label{remark:nested}
    Assumption~\ref{asmptn:decreasing} will automatically hold if the solutions are nested (i.e. $S_{i,j}\subset S_{i,j+1}$) due to submodularity.  Iterative greedy influence maximization methods, such as those based on the  \cite{nemhauser1978analysis}, return nested solutions by design.  Assumption~\ref{asmptn:decreasing} also holds automatically for optimal subsets regardless of nesting (due to submodularity), though it is computationally prohibitive to identify optimal subsets.   
\end{remark}

\subsection{Selecting the final seed set}
After separately solving the influence maximization problem for each community, we allocate the total budget $k$ across the $c$ communities based on the within-community influences $\{\sigma_i(S_{i,j})| i\in\{1,\dots,c\}, j\in\{1,\dots,k\}\}$.Formally, we solve the binary \textit{integer linear program} (ILP) described as Problem~\ref{eq:ILP}.

\begin{problem} \label{eq:ILP}
    \begin{align*}
        & \argmax_{ \{x_{i,j}\}_{\substack{i = 1,\dots,c \\ j = 1,\dots,k}} } \ \sum_{i=1}^{c}\sum_{j=1}^{k} x_{i,j} \sigma_i(S_{i,j}),\\
        \text{s.t.} \ \ \ \ & \sum_{i=1}^{c}\sum_{j=1}^{k} x_{i,j} |S_{i,j}| \le k, \tag{budget constraint} \\
        & \sum_{j=1}^{k} x_{i,j} \le 1 \ \forall i=1,\dots,c, \tag{no repetition}\\
        & x_{i,j} \in \{0,1\} \ \forall i,j.\tag{binary integer constraints}
    \end{align*}
\end{problem}

Before discussing how we propose to solve Problem~\ref{eq:ILP}, we first discuss how we use the solution to this ILP for selecting a seed set and how the objective functions of Problems~\ref{eq:main_problem} and \ref{eq:ILP} relate.

Let $x^*$ denote the optimal solution to Problem~\ref{eq:ILP}.  If there are multiple optimal solutions pick one arbitrarily.  Denote the budget allocated to each community $i$ as $k_i$ (e.g. the index $j$ for which $x^*_{i,j}=1$).    We next construct a seed set for Problem~\ref{eq:main_problem} based on the allocation budget $x^*$,
\begin{align}
    S^{*} \ \gets \ \bigcup_{i=1}^c S_{i,k_i}.
    \label{eq:sol_ilp}
\end{align}

The objective function in Problem~\ref{eq:ILP} lower bounds the objective function of Problem~\ref{eq:main_problem}, with equality if $G$ is formed of disjoint communities.
\begin{theorem}\label{thm:lower-bound}
    Consider any network partition $\{G_i\}_{i=1}^c$ of $G$ and any set of subsets $\{S_i\}_{i=1}^c$ of nodes such that $S_i \subseteq V_i$ for $i=1,\dots,c$.  Then
    \[\sum_{i=1}^c \sigma_i(S_i) \leq \sigma(\cup_{i=1}^c S_i).  \]
\end{theorem}

The proof is in Appendix~\ref*{prf:thm:lower-bound}.

In general, solving an ILP is an NP-Complete problem \cite{kannan1978computational}. However, the submodularity of the influence allows us to solve Problem~\ref{eq:ILP} in polynomial time. 

\subsubsection{Progressive Budgeting} By Assumption~\ref{asmptn:decreasing} (by submodularity for nested subsets), we know that the marginal gain in influence due to each additional node in the seed set is diminishing (both for each individual community and overall since sums of submodular functions are submodular). Hence, we can progressively allocate the budget across the community-based seed sets $\{S_{i,j}\}$. The \text{Progressive-Budgeting} sub-routine used in Algorithm~\ref{alg:community-im} is outlined in Algorithm~\ref{alg:pb}.

\begin{algorithm}
\caption{Progressive-Budgeting}\label{alg:pb}
\begin{algorithmic}[1]
\State \textbf{Input} {$\mathcal{S},\Sigma,k$}.
    \State $\{S_{i,j}| i\in\{1,\dots,c\}, j\in\{1,\dots,k\}\} \gets \mathcal{S}$
    \State $\{\sigma_i(S_{i,j})| i\in\{1,\dots,c\}, j\in\{1,\dots,k\}\} \gets \Sigma$
    \State $\{\delta_i\}_{i=1}^c \gets \{\sigma_i(S_{i,1})\}_{i=1}^c $ \label{line:initdelta} \Comment{Initialize the marginal gains.}
    \State $\{k_i\}_{i=1}^c \gets \{0\}_{i=1}^c \ \forall i$ \label{line:initb} \Comment{Initialize the budget allocations.}
    \State $S^* \gets \emptyset$ \label{line:initsstar} \Comment{Initialize the final set.}
    \For {$\ell = 1,\dots,k$} \label{line:loop}
        \State $m \gets \argmax_{i\in\{1,\dots,c\}}\delta_i$ \label{line:findm} \Comment{Index of the community with the maximum marginal gain.}
        \State $k_m \gets k_m + 1$ \label{line:updateb} \Comment{Update the budget allocated to community $m$.}
        \State $\delta_m \gets \sigma_m(S_{m,{k_m+1}}) - \sigma_m(S_{m,k_m})$ \label{line:updatedelta} \Comment{Update the marginal gains for community $m$.} 
    \EndFor 
    \State $S^* \gets \ \bigcup_{i=1}^c S_{i,k_i}$ 
    \State \textbf{return} $S^{*}$ \Comment{Final seed set.}
\end{algorithmic}
\end{algorithm}

An illustrative example of progressive budgeting is provided in Appendix~\ref*{subsec:pbexample}. We will next discuss the correctness of Algorithm~\ref{alg:pb}. The correctness of Algorithm~\ref{alg:pb} will follow from the following lemma, asserting that up to the uniqueness of optimal solutions of Problem~\ref{eq:ILP} for different cardinalities, the optimal budget allocations are nested.

For each budget $\ell\in\{1,\dots,k\}$, let  $S^{*,(\ell)}$ denote an optimal seed set \eqref{eq:sol_ilp} of cardinality $\ell$ and let $\mathbf{k}^{*,(\ell)} = \{ k_i\}_{i=1}^c$ denote the budget allocations to the $c$ communities.  We say a sequence $(S^{*,(\ell)})_{\ell=1}^k$ of (optimal) seed sets is nested if the seed sets are proper subsets of each other (i.e. $S^{*,(\ell)} \subset S^{*,(\ell+1)}$).  We say a sequence $(\mathbf{k}^{*,(\ell)})_{\ell=1}^k$ of budget allocations is nested if across the sequence each community's allocation is non-decreasing (i.e. $k^{*,(\ell)}_i \leq k^{*,(\ell+1)}_i$).

\begin{lemma} \label{lem:opt-substructure}
 Under Assumption~\ref{asmptn:decreasing}, there is a nested sequence $\{\mathbf{k}^{*,(\ell)} \}_{\ell=1}^k$ of optimal budget allocations for Problem~\ref{eq:ILP}. 
\end{lemma}
The proof is in Appendix~\ref*{prf:lem:opt-substructure}.

\begin{theorem} \label{thm:1}
    Under Assumption~\ref{asmptn:decreasing}, Algorithm~\ref{alg:pb} solves Problem~\ref{eq:ILP}.
\end{theorem}
The proof follows immediately from Lemma~\ref{lem:opt-substructure} and the greedy design of Algorithm~\ref{alg:pb}.

\begin{remark} In general, the guarantees of Theorem~\ref{thm:1} do not translate into guarantees for Problem~\ref{eq:main_problem}.  Since Problem~\ref{eq:main_problem} is an NP-hard problem for common diffusion models on a general network, common methods are approximation algorithms (with an approximation ratio of $(1-1/e)$ or slightly worse) or heuristics.  Thus, the inputs to  Algorithm~\ref{alg:pb} in general will not necessarily be optimal seed sets for their respective communities.  Additionally, as noted in Theorem~\ref{thm:lower-bound}, the objective functions in Problems~\ref{eq:main_problem} and \ref{eq:ILP} only match if the original network $G$ is disjoint (and the communities selected align with the segments of $G$).
\end{remark}

The computational complexity of the proposed framework (Algorithm~\ref{alg:community-im}) is analyzed in Appendix~\ref*{sec:complexity}.
\section{Experiments} \label{section4}
We evaluated the performance of our framework using real-world social networks. We next discuss the network data used for our experiments, list the algorithms chosen for comparison, provide experimental details, and then present results and discussion.

\subsection{Network data}
We used 4 real-world social networks for our experiments. The data is available at \href{https://snap.stanford.edu/data/}{Stanford Large Network Dataset Collection}  \cite{snapnets}. The number of nodes, number of edges, and modularity (for the network-partition obtained using the Louvain method \cite{blondel2008fast}) of each network are provided in Table~\ref{tab:basic_info}.

\begin{table}[h!] 
	\caption{Basic information of the networks used.} \label{tab:basic_info}
	\centering
	\begin{tabular}{l r r c}
		\hline \hline
		Network & Nodes & Edges & Modularity\\
		\hline
		Facebook \cite{leskovec2012learning} & 4,039 & 88,234 & 0.8678 \\
		Bitcoin \cite{kumar2016edge, kumar2018rev2} & 5,881 & 35,592 & 0.4196 \\
		Wikipedia \cite{leskovec2010signed, leskovec2010predicting} & 7,115 & 103,689 & 0.4175 \\
		Epinions \cite{richardson2003trust} & 75,879 & 508,837 & 0.8219 \\
		\hline
	\end{tabular}
\end{table}

The Facebook network \cite{leskovec2012learning} consists of a dataset consisting of `circles' (or `friends lists') from Facebook. The Facebook network is undirected; we converted it to a directed network by replacing each edge with two directed edges. Bitcoin  network \cite{kumar2016edge, kumar2018rev2} is a (directed) who-trusts-whom network of people who trade using Bitcoin on a platform called Bitcoin OTC. Wikipedia network \cite{leskovec2010signed, leskovec2010predicting} is a who-votes-on-whom (directed) network to become an administrator. Epinions network \cite{richardson2003trust} is a who-trust-whom (directed) online social network of a general consumer review platform called Epinions.

For edge-weights, two models are used which are \textit{weighted cascade} (WC) model \cite{kempe2003maximizing} where for each node $v\in V$, the weight of each edge entering $v$ was set to $1/\text{in-degree}(v)$ and \textit{trivalency} (TV) model \cite{goyal2011data} where each edge-weight was drawn uniformly at random from a small set of constants \{0.1, 0.01, 0.001\}. However, for the linear threshold model (LT) of diffusion, only the WC model is used for edge-weights as the TV model does not necessarily maintain the sum of weights of all edges incident on a node to be less than or equal to 1.

\subsection{Algorithms} \label{exp:algos} 
We compared the proposed community-aware framework (\text{Community-IM}) with the following algorithms. 
\begin{enumerate}
    \item \text{CELF++} \cite{goyal2011celf++}, the state-of-the-art simulation-based greedy algorithm. 
    \item \text{CoFIM} \cite{shang2017cofim}, a community-aware heuristic algorithm with guarantees under the independent cascade diffusion model with the weighted-cascade edge-weight model.
    \item \text{DSGA} \cite{lotf2022improved}, a recent genetic algorithm-based method that uses Monte Carlo simulations.
    \item \text{Degree} \cite{kempe2003maximizing}, the simplest heuristic algorithm where for budget $k$, top-$k$ out-degree nodes are selected.
    \item \text{Degree-Discount} \cite{chen2009efficient}, a modification of the the \text{Degree} heuristic algorithm with better empirical performance.
\end{enumerate}

Note that the \text{CoFIM} algorithm was developed only for IC diffusion model with WC edge-weight model. However, for empirical comparisons, we implemented it for the other choices of diffusion models and edge-weight models as well. 

For the purpose of demonstrating the performance of the proposed framework, \text{Community-IM} (Algorithm~\ref{alg:community-im}), we used the Louvain method \cite{blondel2008fast} as \texttt{com-method}, and \text{CELF++} \cite{goyal2011celf++} as \texttt{sol-method} for \text{Community-Detection} and \text{Generate-Candidates} subroutines, respectively. In general, the user may try different combinations of \texttt{com-method} and \texttt{sol-method} as part of the proposed framework.

We also studied the effect of the (modularity of) the community structure on the performance of the proposed framework. We used the Louvain  \cite{blondel2008fast},  Label Propagation \cite{cordasco2010community}, and  Girvan-Newman  \cite{girvan2002community} community-detection methods (discussed in Section~\ref{sec:community-detection-methods}) as \texttt{com-method} in Community-Detection step of the proposed framework. For brevity, we only considered the Facebook network under different diffusion models and WC edge-weight model.

\subsection{Experimental details} \label{exp:details} 
We used the budgets $k=1,5,10,\dots,100$ for comparing different algorithms. However, for DSGA \cite{lotf2022improved}, we only used the budgets $k = 1,20,40,\dots,100$ due to its high run-time. For brevity, for studying the effect of the community structure on the proposed framework, we used the budgets $k=1,5,\dots,50$. The influence of any seed set was estimated as the average number of active nodes from $1,000$ different Monte Carlo simulations of the underlying diffusion starting with the same seed set. For any network, if a community detection method returned some communities whose individual sizes are below 1\% of the number of nodes in the network then we merged them all into a single community. We do this to avoid having too many small communities.

The experiments were carried out on a computer with 2.6 GHz 24-core Intel Xeon Gold Sky Lake processors and 96 GB of memory. We used Python for our implementation. The source codes of CELF++ and CoFIM provided by their authors are written in C++. The data and source code for this paper are available \href{https://github.com/abhishekumrawal/Community-IM}{here}. 

\subsection{Results} 
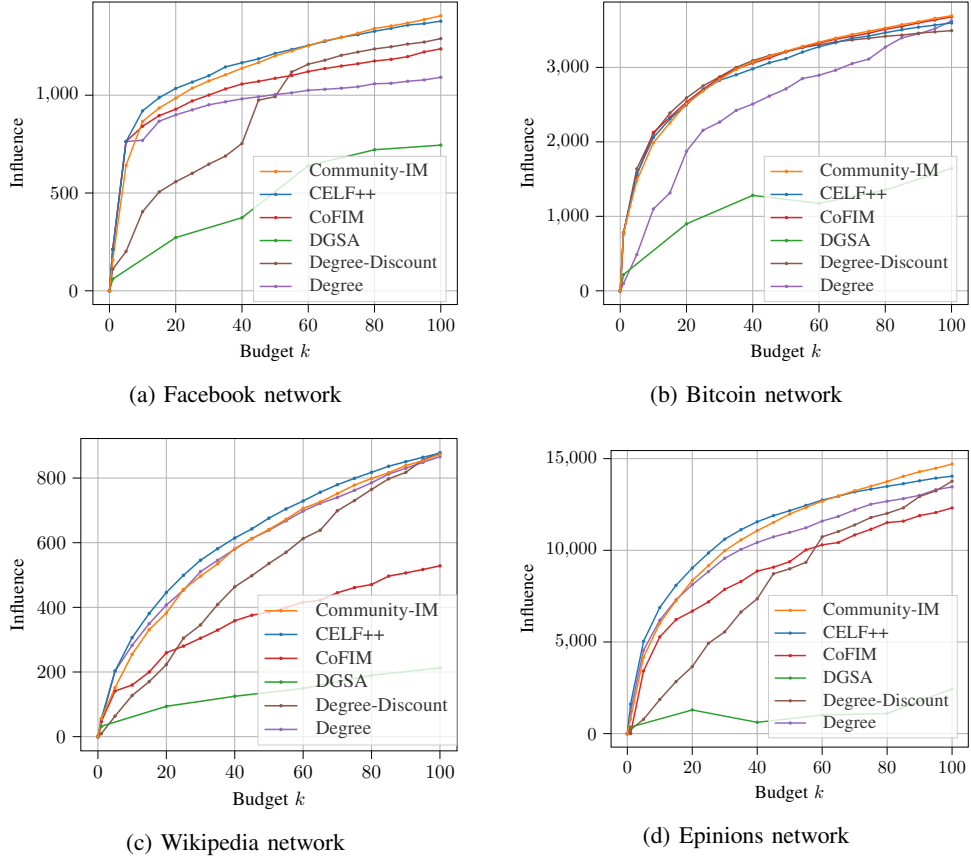
\begin{figure*}[t] 
\centering
\begin{minipage}{.34\textwidth}
  \centering
  \resizebox{\textwidth}{!}{
\begin{tikzpicture}

\definecolor{color0}{rgb}{0.12156862745098,0.466666666666667,0.705882352941177}
\definecolor{color1}{rgb}{1,0.498039215686275,0.0549019607843137}
\definecolor{color2}{rgb}{0.172549019607843,0.627450980392157,0.172549019607843}
\definecolor{color3}{rgb}{0.83921568627451,0.152941176470588,0.156862745098039}
\definecolor{color4}{rgb}{0.580392156862745,0.403921568627451,0.741176470588235}
\definecolor{color5}{rgb}{0.549019607843137,0.337254901960784,0.294117647058824}

\begin{axis}[
legend cell align={left},
reverse legend,
legend style={
  fill opacity=0.8,
  draw opacity=1,
  text opacity=1,
  at={(0.97,0.25)},
  anchor=east,
  draw=white!80!black
},
tick align=outside,
tick pos=left,
x grid style={white!69.0196078431373!black},
xlabel={Budget \(\displaystyle k\)},
xmajorgrids,
xmin=-5, xmax=105,
xtick style={color=black},
y grid style={white!69.0196078431373!black},
ylabel={Influence},
ymajorgrids,
ymin=-70.30085, ymax=1476.31785,
ytick style={color=black}
]

\addplot [thick, color4, mark=*, mark size=.5, mark options={solid}]
table {%
0 0
1 211.591003417969
5 763.094970703125
10 769.320983886719
15 867.013977050781
20 899.322998046875
25 925.156005859375
30 950.905029296875
35 966.216979980469
40 981.385009765625
45 991.741027832031
50 1003.36297607422
55 1012.19500732422
60 1025.1669921875
65 1030.12194824219
70 1035.67602539062
75 1043.13098144531
80 1058.4990234375
85 1060.81799316406
90 1071.20300292969
95 1078.13098144531
100 1091.65405273438
};
\addlegendentry{Degree}
\addplot [thick, color5, mark=*, mark size=.5, mark options={solid}]
table {%
0 0
1 110.985000610352
5 201.10400390625
10 404.919006347656
15 505.459014892578
20 557.736999511719
25 600.783020019531
30 647.387023925781
35 688.68701171875
40 753.315979003906
45 974.864990234375
50 992.361022949219
55 1118.78295898438
60 1158.38903808594
65 1178.96801757812
70 1203.96203613281
75 1220.78002929688
80 1236.56396484375
85 1247.5830078125
90 1261.26293945312
95 1271.63305664062
100 1289.28002929688
};
\addlegendentry{Degree-Discount}
\addplot [thick, color2, mark=*, mark size=.5, mark options={solid}]
table {%
0 0
1 60.4900016784668
20 272.209991455078
40 373.727996826172
60 639.5419921875
80 721.166015625
100 744.75
};
\addlegendentry{DGSA}
\addplot [thick, color3, mark=*, mark size=.5, mark options={solid}]
table {%
0 0
1 211.591003417969
5 763.094970703125
10 841.512023925781
15 895.614990234375
20 928.309020996094
25 970.783020019531
30 1000.71997070312
35 1032
40 1057.13000488281
45 1070.77795410156
50 1086.30700683594
55 1100.82604980469
60 1121.09497070312
65 1136.39196777344
70 1149.15002441406
75 1160.80395507812
80 1174.89697265625
85 1183.31396484375
90 1197.06396484375
95 1221.44299316406
100 1236.77600097656
};
\addlegendentry{CoFIM}
\addplot [thick, color0, mark=*, mark size=.5, mark options={solid}]
table {%
0 0
1 211.591003417969
5 763.094970703125
10 920.122985839844
15 987.211975097656
20 1034.61999511719
25 1067.23999023438
30 1099.81005859375
35 1144.47998046875
40 1165.90002441406
45 1185.94995117188
50 1213.57995605469
55 1233.81005859375
60 1254.44995117188
65 1274.58996582031
70 1296.48999023438
75 1309.4599609375
80 1327.03002929688
85 1341.81994628906
90 1358.34997558594
95 1366.2099609375
100 1378.39001464844
};
\addlegendentry{CELF++}
\addplot [thick, color1, mark=*, mark size=.5, mark options={solid}]
table {%
0 0
1 157.576995849609
5 642.02001953125
10 865.966003417969
15 934.627014160156
20 984.731018066406
25 1036.02001953125
30 1072.7919921875
35 1104.10803222656
40 1138.00305175781
45 1166.71899414062
50 1201.19897460938
55 1226.34301757812
60 1252.32202148438
65 1278.791015625
70 1295.39794921875
75 1317.05700683594
80 1341.23400878906
85 1354.15295410156
90 1368.84301757812
95 1386.83605957031
100 1406.01696777344
};
\addlegendentry{Community-IM}
\end{axis}

\end{tikzpicture}}
  \subcaption{Facebook network}
\end{minipage}\hspace{.5cm}
\vspace{.15in}
\begin{minipage}{.34\textwidth}
  \centering
  \resizebox{\textwidth}{!}{
\begin{tikzpicture}

\definecolor{color0}{rgb}{0.12156862745098,0.466666666666667,0.705882352941177}
\definecolor{color1}{rgb}{1,0.498039215686275,0.0549019607843137}
\definecolor{color2}{rgb}{0.172549019607843,0.627450980392157,0.172549019607843}
\definecolor{color3}{rgb}{0.83921568627451,0.152941176470588,0.156862745098039}
\definecolor{color4}{rgb}{0.580392156862745,0.403921568627451,0.741176470588235}
\definecolor{color5}{rgb}{0.549019607843137,0.337254901960784,0.294117647058824}

\begin{axis}[
legend cell align={left},
reverse legend,
legend style={
  fill opacity=0.8,
  draw opacity=1,
  text opacity=1,
  at={(0.97,0.25)},
  anchor=east,
  draw=white!80!black
},
tick align=outside,
tick pos=left,
x grid style={white!69.0196078431373!black},
xlabel={Budget \(\displaystyle k\)},
xmajorgrids,
xmin=-5, xmax=105,
xtick style={color=black},
y grid style={white!69.0196078431373!black},
ylabel={Influence},
ymajorgrids,
ymin=-184.6443, ymax=3877.5303,
ytick style={color=black}
]

\addlegendentry{Degree}
\addplot [thick, color4, mark=*, mark size=.5, mark options={solid}]
table {%
0 0
1 98.6070022583008
5 486.110992431641
10 1098.83203125
15 1314.94104003906
20 1873.07800292969
25 2155.57202148438
30 2265.13891601562
35 2422.48388671875
40 2507.8701171875
45 2614.43188476562
50 2710.97705078125
55 2850.212890625
60 2894.30102539062
65 2962.2109375
70 3050.89111328125
75 3111.49389648438
80 3271.580078125
85 3397.17993164062
90 3450.04907226562
95 3519.48999023438
100 3625.05810546875
};
\addlegendentry{Degree-Discount}
\addplot [thick, color5, mark=*, mark size=.5, mark options={solid}]
table {%
0 0
1 778.656982421875
5 1638.47998046875
10 2103.72998046875
15 2387.2900390625
20 2589.6201171875
25 2751.19995117188
30 2873.03002929688
35 2999.51000976562
40 3091.02001953125
45 3158.56005859375
50 3212.1201171875
55 3266.28002929688
60 3303.4599609375
65 3341.30004882812
70 3368.19995117188
75 3391.17993164062
80 3417.78002929688
85 3433.1298828125
90 3458.9599609375
95 3477.419921875
100 3493.25
};
\addplot [thick, color2, mark=*, mark size=.5, mark options={solid}]
table {%
0 0
1 217.912994384766
20 898.843017578125
40 1281.91003417969
60 1174.07299804688
80 1353.72802734375
100 1643.26599121094
};
\addlegendentry{DGSA}
\addplot [thick, color3, mark=*, mark size=.5, mark options={solid}]
table {%
0 0
1 778.656982421875
5 1551.9189453125
10 2126.27001953125
15 2331.69311523438
20 2533.72900390625
25 2705.9208984375
30 2860.76391601562
35 2978.32006835938
40 3058.27709960938
45 3126.36108398438
50 3208.123046875
55 3272.25805664062
60 3326.7080078125
65 3378.38696289062
70 3417.32299804688
75 3458.49096679688
80 3511.85595703125
85 3549.21997070312
90 3597.03491210938
95 3637.22094726562
100 3678.58203125
};
\addlegendentry{CoFIM}
\addplot [thick, color0, mark=*, mark size=.5, mark options={solid}]
table {%
0 0
1 778.656982421875
5 1571.04699707031
10 2055.83203125
15 2309
20 2496.955078125
25 2698.38891601562
30 2826.86791992188
35 2900.68701171875
40 2980.27197265625
45 3061.94189453125
50 3117.52099609375
55 3205.294921875
60 3276.82690429688
65 3331.73193359375
70 3390.44506835938
75 3420.62109375
80 3465.78002929688
85 3505.43408203125
90 3539.94604492188
95 3567.93701171875
100 3596.40405273438
};
\addlegendentry{CELF++}
\addplot [thick, color1, mark=*, mark size=.5, mark options={solid}]
table {%
0 0
1 778.656982421875
5 1479.65600585938
10 1988.09399414062
15 2256.34008789062
20 2509.81201171875
25 2682.59301757812
30 2838.5439453125
35 2972.27905273438
40 3071.97412109375
45 3149.73901367188
50 3218.27392578125
55 3280.7099609375
60 3338.787109375
65 3392.9140625
70 3441.74389648438
75 3485.70288085938
80 3528.36889648438
85 3573.85791015625
90 3612.55908203125
95 3658.29809570312
100 3692.88598632812
};
\addlegendentry{Community-IM}
\end{axis}

\end{tikzpicture}}
  \subcaption{Bitcoin network}
\end{minipage}
\begin{minipage}{.34\textwidth}
  \centering
  \resizebox{\textwidth}{!}{
\begin{tikzpicture}

\definecolor{color0}{rgb}{0.12156862745098,0.466666666666667,0.705882352941177}
\definecolor{color1}{rgb}{1,0.498039215686275,0.0549019607843137}
\definecolor{color2}{rgb}{0.172549019607843,0.627450980392157,0.172549019607843}
\definecolor{color3}{rgb}{0.83921568627451,0.152941176470588,0.156862745098039}
\definecolor{color4}{rgb}{0.580392156862745,0.403921568627451,0.741176470588235}
\definecolor{color5}{rgb}{0.549019607843137,0.337254901960784,0.294117647058824}

\begin{axis}[
legend cell align={left},
reverse legend,
legend style={
  fill opacity=0.8,
  draw opacity=1,
  text opacity=1,
  at={(0.47,0.5)},
  anchor=north west,
  draw=white!80!black
},
tick align=outside,
tick pos=left,
x grid style={white!69.0196078431373!black},
xlabel={Budget \(\displaystyle k\)},
xmajorgrids,
xmin=-5, xmax=105,
xtick style={color=black},
y grid style={white!69.0196078431373!black},
ylabel={Influence},
ymajorgrids,
ymin=-43.9182, ymax=922.2822,
ytick style={color=black}
]

\addplot [thick, color4, mark=*, mark size=.5, mark options={solid}]
table {%
0 0
1 47.0139999389648
5 202.625
10 282.484985351562
15 349.643005371094
20 407.055999755859
25 453.342010498047
30 510.497985839844
35 545.596008300781
40 580.072021484375
45 612.806030273438
50 638.752014160156
55 667.718994140625
60 697.955993652344
65 721.875
70 739.828979492188
75 761.9990234375
80 785.335998535156
85 812.276000976562
90 830.18798828125
95 848.557983398438
100 866.060974121094
};
\addlegendentry{Degree}
\addplot [thick, color5, mark=*, mark size=.5, mark options={solid}]
table {%
0 0
1 9.22000026702881
5 63.5
10 127.346000671387
15 170.272994995117
20 222.822006225586
25 304.834991455078
30 345.558990478516
35 408.813995361328
40 463.429992675781
45 498.117004394531
50 535.705017089844
55 570.327026367188
60 612.780029296875
65 638.692016601562
70 698.844970703125
75 730.317016601562
80 764.843994140625
85 797.661010742188
90 817.534973144531
95 856.320983886719
100 878.364013671875
};
\addlegendentry{Degree-Discount}
\addplot [thick, color2, mark=*, mark size=.5, mark options={solid}]
table {%
0 0
1 32.1059989929199
20 93.7720031738281
40 124.946998596191
60 149.162002563477
80 189.815002441406
100 212.60400390625
};
\addlegendentry{DGSA}
\addplot [thick, color3, mark=*, mark size=.5, mark options={solid}]
table {%
0 0
1 47.0139999389648
5 141.095001220703
10 159.764999389648
15 199.977996826172
20 259.399993896484
25 280.100006103516
30 304.622985839844
35 329.572998046875
40 358.552001953125
45 375.634002685547
50 385.078002929688
55 399.238006591797
60 415.337005615234
65 422.023986816406
70 445.138000488281
75 461.054992675781
80 470.64599609375
85 496.631011962891
90 506.472991943359
95 517.117004394531
100 528.357971191406
};
\addlegendentry{CoFIM}
\addplot [thick, color0, mark=*, mark size=.5, mark options={solid}]
table {%
0 0
1 55.2019996643066
5 202.625
10 306.363006591797
15 381.265014648438
20 446.032012939453
25 499.177001953125
30 545.828002929688
35 581.489013671875
40 614.598999023438
45 642.864990234375
50 676.012023925781
55 704.432983398438
60 729.1669921875
65 756.142028808594
70 779.328002929688
75 799.112976074219
80 817.768981933594
85 836.572021484375
90 850.619995117188
95 864.232971191406
100 876.906005859375
};
\addlegendentry{CELF++}
\addplot [thick, color1, mark=*, mark size=.5, mark options={solid}]
table {%
0 0
1 55.2019996643066
5 151.485000610352
10 254.690002441406
15 331.147003173828
20 382.480010986328
25 455.227996826172
30 496.441986083984
35 534.68798828125
40 581.716003417969
45 612.476013183594
50 641.135986328125
55 672.367004394531
60 706.473022460938
65 726.119995117188
70 751.869018554688
75 778.205993652344
80 798.72900390625
85 816.460021972656
90 838.661010742188
95 853.575988769531
100 873.043029785156
};
\addlegendentry{Community-IM}
\end{axis}

\end{tikzpicture}}
  \subcaption{Wikipedia network}
\end{minipage}\hspace{.5cm}
\begin{minipage}{.34\textwidth}
  \centering
  \resizebox{\textwidth}{!}{
\begin{tikzpicture}

\definecolor{color0}{rgb}{0.12156862745098,0.466666666666667,0.705882352941177}
\definecolor{color1}{rgb}{1,0.498039215686275,0.0549019607843137}
\definecolor{color2}{rgb}{0.172549019607843,0.627450980392157,0.172549019607843}
\definecolor{color3}{rgb}{0.83921568627451,0.152941176470588,0.156862745098039}
\definecolor{color4}{rgb}{0.580392156862745,0.403921568627451,0.741176470588235}
\definecolor{color5}{rgb}{0.549019607843137,0.337254901960784,0.294117647058824}

\begin{axis}[
legend cell align={left},
reverse legend,
legend style={fill opacity=0.8, draw opacity=1, text opacity=1, at={(0.97,0.03)}, anchor=south east, draw=white!80!black},
tick align=outside,
tick pos=left,
x grid style={white!69.0196078431373!black},
xlabel={Budget \(\displaystyle k\)},
xmajorgrids,
xmin=-5, xmax=105,
xtick style={color=black},
y grid style={white!69.0196078431373!black},
ylabel={Influence},
ymajorgrids,
ymin=-735.30365, ymax=15441.37665,
ytick style={color=black}
]

\addplot [thick, color4, mark=*, mark size=.5, mark options={solid}]
table {%
0 0
1 1215.623046875
5 4545.72119140625
10 6189.1181640625
15 7300.63818359375
20 8121.56396484375
25 8842.2392578125
30 9566.25
35 10048.505859375
40 10425.37109375
45 10732.9833984375
50 10975.0546875
55 11227.6953125
60 11593.896484375
65 11850.1630859375
70 12203.853515625
75 12514.388671875
80 12674.2197265625
85 12819.2021484375
90 12998.736328125
95 13307.4287109375
100 13457.5087890625
};
\addlegendentry{Degree}
\addplot [thick, color5, mark=*, mark size=.5, mark options={solid}]
table {%
0 0
1 215.891006469727
5 781.526977539062
10 1860.92199707031
15 2836.8310546875
20 3655.69702148438
25 4930.52880859375
30 5551.60986328125
35 6633.158203125
40 7355.23388671875
45 8718.2880859375
50 8989.9013671875
55 9344.8564453125
60 10734.720703125
65 11027.646484375
70 11382.6328125
75 11788.8154296875
80 12016.4208984375
85 12315.673828125
90 12936.5390625
95 13242.1962890625
100 13771.251953125
};
\addlegendentry{Degree-Discount}
\addplot [thick, color2, mark=*, mark size=.5, mark options={solid}]
table {%
0 0
1 357.575012207031
20 1292.36499023438
40 616.715026855469
60 1015.43200683594
80 1097.15795898438
100 2439.23608398438
};
\addlegendentry{DGSA}
\addplot [thick, color3, mark=*, mark size=.5, mark options={solid}]
table {%
0 0
1 9.42899990081787
5 3419.18090820312
10 5277.61376953125
15 6218.72412109375
20 6682.04296875
25 7189.14990234375
30 7870.72998046875
35 8298.595703125
40 8866.802734375
45 9077.240234375
50 9379.6435546875
55 10022.9658203125
60 10297.4189453125
65 10424.9052734375
70 10837.50390625
75 11149.6279296875
80 11510.443359375
85 11589.7236328125
90 11893.205078125
95 12061.259765625
100 12314.9658203125
};
\addlegendentry{CoFIM}
\addplot [thick, color0, mark=*, mark size=.5, mark options={solid}]
table {%
0 0
1 1612.18994140625
5 5035.02001953125
10 6882.4501953125
15 8089.2001953125
20 9032.8896484375
25 9855.9501953125
30 10608.400390625
35 11124.2998046875
40 11556.099609375
45 11892.7998046875
50 12159.099609375
55 12452.099609375
60 12742.099609375
65 12950.5
70 13182.2998046875
75 13336.900390625
80 13490.599609375
85 13633.5
90 13791.400390625
95 13938.400390625
100 14042.599609375
};
\addlegendentry{CELF++}
\addplot [thick, color1, mark=*, mark size=.5, mark options={solid}]
table {%
0 0
1 787.606018066406
5 4173.7490234375
10 6027.19189453125
15 7248.76123046875
20 8355.724609375
25 9166.345703125
30 9978.7177734375
35 10565.8505859375
40 11072.3330078125
45 11519.646484375
50 11980.48828125
55 12339.5439453125
60 12681.1552734375
65 12964.30078125
70 13247.908203125
75 13496.0244140625
80 13753.8154296875
85 14034.71484375
90 14287.6171875
95 14476.533203125
100 14706.0732421875
};
\addlegendentry{Community-IM}
\end{axis}

\end{tikzpicture}}
  \subcaption{Epinions network}
\end{minipage}
\caption{Influence vs. budget $k$ for different networks under IC diffusion model and WC edge-weight model.} \label{fig:ic-wc}
\end{figure*}

\begin{figure*}[t]
\centering
\begin{minipage}{.33\textwidth}
  \centering
  \resizebox{\textwidth}{!}{
\begin{tikzpicture}

\definecolor{color0}{rgb}{0.12156862745098,0.466666666666667,0.705882352941177}
\definecolor{color1}{rgb}{1,0.498039215686275,0.0549019607843137}
\definecolor{color2}{rgb}{0.172549019607843,0.627450980392157,0.172549019607843}
\definecolor{color3}{rgb}{0.83921568627451,0.152941176470588,0.156862745098039}
\definecolor{color4}{rgb}{0.580392156862745,0.403921568627451,0.741176470588235}
\definecolor{color5}{rgb}{0.549019607843137,0.337254901960784,0.294117647058824}

\begin{axis}[
legend cell align={left},
reverse legend,
legend style={
  fill opacity=0.8,
  draw opacity=1,
  text opacity=1,
  at={(0.97,0.01)},
  anchor=south east,
  draw=white!80!black
},
tick align=outside,
tick pos=left,
x grid style={white!69.0196078431373!black},
xlabel={Budget \(\displaystyle k\)},
xmajorgrids,
xmin=-5, xmax=105,
xtick style={color=black},
y grid style={white!69.0196078431373!black},
ylabel={Influence},
ymajorgrids,
ymin=-98.87095, ymax=2076.28995,
ytick style={color=black}
]

\addplot [thick, color4, mark=*, mark size=.5, mark options={solid}]
table {%
0 0
1 1335.04504394531
5 1764.28002929688
10 1763.97802734375
15 1765.72302246094
20 1763.64599609375
25 1763.47094726562
30 1764.45300292969
35 1764.42895507812
40 1763.88000488281
45 1766.41101074219
50 1766.49499511719
55 1765.98400878906
60 1762.22399902344
65 1765.29699707031
70 1765.98999023438
75 1763.99401855469
80 1767.55505371094
85 1765.2080078125
90 1766.51196289062
95 1765.21594238281
100 1765.08703613281
};
\addlegendentry{Degree}
\addplot [thick, color5, mark=*, mark size=.5, mark options={solid}]
table {%
0 0
1 399.848999023438
5 1454.84704589844
10 1758.5810546875
15 1793.25
20 1794.73803710938
25 1797.03002929688
30 1797.95495605469
35 1799.83605957031
40 1796.50695800781
45 1802.17297363281
50 1798.94702148438
55 1799.99401855469
60 1799.37194824219
65 1797.18798828125
70 1798.97595214844
75 1802.04699707031
80 1798.16198730469
85 1802.5400390625
90 1801.93896484375
95 1802.2900390625
100 1801.25598144531
};
\addlegendentry{Degree-Discount}
\addplot [thick, color2, mark=*, mark size=.5, mark options={solid}]
table {%
0 0
1 1310.58898925781
20 1711.06994628906
40 1760.38305664062
60 1716.03601074219
80 1781.09802246094
100 1784.73498535156
};
\addlegendentry{DGSA}
\addplot [thick, color3, mark=*, mark size=.5, mark options={solid}]
table {%
0 0
1 1335.04504394531
5 1764.28002929688
10 1803.8349609375
15 1804.04797363281
20 1803.47204589844
25 1802.80102539062
30 1802.47900390625
35 1804.64196777344
40 1802.85095214844
45 1807.81799316406
50 1805.22497558594
55 1806.05603027344
60 1805.55798339844
65 1805.02502441406
70 1806.01403808594
75 1805.55603027344
80 1809.53698730469
85 1805.16296386719
90 1807.42297363281
95 1809.82702636719
100 1809.11901855469
};
\addlegendentry{CoFIM}
\addplot [thick, color0, mark=*, mark size=.5, mark options={solid}]
table {%
0 0
1 1387.44995117188
5 1791.82995605469
10 1851.76000976562
15 1859.46997070312
20 1866.93005371094
25 1878.21997070312
30 1883.0400390625
35 1891.08996582031
40 1901.44995117188
45 1905.22998046875
50 1916.5400390625
55 1924.32995605469
60 1928.64001464844
65 1937.11999511719
70 1941.43994140625
75 1946.77001953125
80 1956.69995117188
85 1962.21997070312
90 1967.31994628906
95 1969.47998046875
100 1977.27001953125
};
\addlegendentry{CELF++}
\addplot [thick, color1, mark=*, mark size=.5, mark options={solid}]
table {%
0 0
1 1323.79296875
5 1643.59204101562
10 1766.17199707031
15 1817.11303710938
20 1834.64697265625
25 1849.26501464844
30 1859.23400878906
35 1872.84802246094
40 1882.32995605469
45 1892.58801269531
50 1899.93200683594
55 1912.35803222656
60 1919.52502441406
65 1927.12805175781
70 1937.3740234375
75 1948.26501464844
80 1951.5810546875
85 1960.91296386719
90 1964.0419921875
95 1972.38305664062
100 1977.4189453125
};
\addlegendentry{Community-IM}
\end{axis}

\end{tikzpicture}}
  \subcaption{Facebook network}
\end{minipage}
\begin{minipage}{.33\textwidth}
  \centering
  \resizebox{\textwidth}{!}{
\begin{tikzpicture}

\definecolor{color0}{rgb}{0.12156862745098,0.466666666666667,0.705882352941177}
\definecolor{color1}{rgb}{1,0.498039215686275,0.0549019607843137}
\definecolor{color2}{rgb}{0.172549019607843,0.627450980392157,0.172549019607843}
\definecolor{color3}{rgb}{0.83921568627451,0.152941176470588,0.156862745098039}
\definecolor{color4}{rgb}{0.580392156862745,0.403921568627451,0.741176470588235}
\definecolor{color5}{rgb}{0.549019607843137,0.337254901960784,0.294117647058824}

\begin{axis}[
legend cell align={left},
reverse legend,
legend style={
  fill opacity=0.8,
  draw opacity=1,
  text opacity=1,
  at={(0.97,0.25)},
  anchor=east,
  draw=white!80!black
},
tick align=outside,
tick pos=left,
x grid style={white!69.0196078431373!black},
xlabel={Budget \(\displaystyle k\)},
xmajorgrids,
xmin=-5, xmax=105,
xtick style={color=black},
y grid style={white!69.0196078431373!black},
ylabel={Influence},
ymajorgrids,
ymin=-28.1137, ymax=590.3877,
ytick style={color=black}
]
\addplot [thick, color0, mark=*, mark size=.5, mark options={solid}]
table {%
0 0
1 347.470001220703
5 398.622985839844
10 410.019989013672
15 421.100006103516
20 429.841003417969
25 436.135009765625
30 444.368988037109
35 451.752014160156
40 462.106994628906
45 469.0419921875
50 478.644012451172
55 486.110992431641
60 494.751007080078
65 503.618011474609
70 509.680999755859
75 518.1240234375
80 524.4169921875
85 531.307983398438
90 538.06201171875
95 544.869018554688
100 550.901000976562
};
\addlegendentry{CELF++}
\addplot [thick, color3, mark=*, mark size=.5, mark options={solid}]
table {%
0 0
1 314.710998535156
5 365.316986083984
10 390.114013671875
15 402.743988037109
20 420.015991210938
25 430.296997070312
30 440.037994384766
35 451.457000732422
40 458.601013183594
45 467.618011474609
50 479.493011474609
55 485.188995361328
60 490.811004638672
65 502.042999267578
70 507.200988769531
75 517.393005371094
80 524.218017578125
85 531.7080078125
90 539.133972167969
95 544.070007324219
100 550.689025878906
};
\addlegendentry{CoFIM}
\addplot [thick, color2, mark=*, mark size=.5, mark options={solid}]
table {%
0 0
1 317.449005126953
20 369.674987792969
40 397.092010498047
60 439.860992431641
80 459.170013427734
100 487.337005615234
};
\addlegendentry{DGSA}
\addplot [thick, color4, mark=*, mark size=.5, mark options={solid}]
table {%
0 0
1 314.710998535156
5 365.726013183594
10 386.135009765625
15 394.35400390625
20 408.911010742188
25 419.321990966797
30 432.691986083984
35 439.457000732422
40 444.036010742188
45 455.796997070312
50 461.174011230469
55 470.684997558594
60 479.205993652344
65 485.537994384766
70 492.904998779297
75 498.497985839844
80 506.466003417969
85 514.85302734375
90 519.960998535156
95 526.176025390625
100 531.739990234375
};
\addlegendentry{Degree}
\addplot [thick, color5, mark=*, mark size=.5, mark options={solid}]
table {%
0 0
1 71.6190032958984
5 340.075012207031
10 379.683013916016
15 384.877990722656
20 396.054992675781
25 411.554992675781
30 423.536010742188
35 435.864990234375
40 444.601013183594
45 451.908996582031
50 460.144012451172
55 472.113006591797
60 480.871002197266
65 488.972991943359
70 499.403015136719
75 506.002990722656
80 511.337005615234
85 520.906005859375
90 533.182983398438
95 541.60302734375
100 548.236999511719
};
\addlegendentry{Degree-Discount}
\addplot [thick, color1, mark=*, mark size=.5, mark options={solid}]
table {%
0 0
1 317.449005126953
5 368.834991455078
10 392.914001464844
15 405.740997314453
20 416.207000732422
25 430.542999267578
30 442.993011474609
35 457.855987548828
40 466.601989746094
45 475.381988525391
50 482.601013183594
55 491.652008056641
60 501.1669921875
65 510.114990234375
70 518.398986816406
75 526.72802734375
80 534.723022460938
85 539.815002441406
90 549.908996582031
95 555.723999023438
100 562.273986816406
};
\addlegendentry{Community-IM}
\end{axis}

\end{tikzpicture}}
  \subcaption{Bitcoin network}
\end{minipage}
\begin{minipage}{.33\textwidth}
  \centering
  \resizebox{\textwidth}{!}{
\begin{tikzpicture}

\definecolor{color0}{rgb}{0.12156862745098,0.466666666666667,0.705882352941177}
\definecolor{color1}{rgb}{1,0.498039215686275,0.0549019607843137}
\definecolor{color2}{rgb}{0.172549019607843,0.627450980392157,0.172549019607843}
\definecolor{color3}{rgb}{0.83921568627451,0.152941176470588,0.156862745098039}
\definecolor{color4}{rgb}{0.580392156862745,0.403921568627451,0.741176470588235}
\definecolor{color5}{rgb}{0.549019607843137,0.337254901960784,0.294117647058824}

\begin{axis}[
legend cell align={left},
reverse legend,
legend style={
  fill opacity=0.8,
  draw opacity=1,
  text opacity=1,
  at={(0.97,0.26)},
  anchor=east,
  draw=white!80!black
},
tick align=outside,
tick pos=left,
x grid style={white!69.0196078431373!black},
xlabel={Budget \(\displaystyle k\)},
xmajorgrids,
xmin=-5, xmax=105,
xtick style={color=black},
y grid style={white!69.0196078431373!black},
ylabel={Influence},
ymajorgrids,
ymin=-61.7615, ymax=1296.9915,
ytick style={color=black}
]

\addplot [thick, color4, mark=*, mark size=.5, mark options={solid}]
table {%
0 0
1 685.265991210938
5 822.705017089844
10 852.565979003906
15 878.492980957031
20 917.640014648438
25 942.380981445312
30 983.515991210938
35 993.617980957031
40 1003.625
45 1020.11602783203
50 1031.8759765625
55 1039.67004394531
60 1067.21801757812
65 1077.53405761719
70 1088.55700683594
75 1105.51293945312
80 1123.80004882812
85 1134.82495117188
90 1143.77099609375
95 1149.45397949219
100 1152.00598144531
};
\addlegendentry{Degree}
\addplot [thick, color5, mark=*, mark size=.5, mark options={solid}]
table {%
0 0
1 625.768005371094
5 734.781005859375
10 773.526000976562
15 789.921997070312
20 814.247009277344
25 864.515991210938
30 896.736999511719
35 936.723022460938
40 948.453979492188
45 976.724975585938
50 1003.83197021484
55 1017.82800292969
60 1044.42700195312
65 1093.208984375
70 1102.89294433594
75 1118.92199707031
80 1129.90002441406
85 1145.97998046875
90 1163.34899902344
95 1170.13903808594
100 1183.10900878906
};
\addlegendentry{Degree-Discount}
\addplot [thick, color2, mark=*, mark size=.5, mark options={solid}]
table {%
0 0
1 665.151000976562
20 716.484985351562
40 757.85302734375
60 788.390991210938
80 821.546020507812
100 847.632995605469
};
\addlegendentry{DGSA}
\addplot [thick, color3, mark=*, mark size=.5, mark options={solid}]
table {%
0 0
1 685.265991210938
5 775.4169921875
10 782.945007324219
15 790.138000488281
20 810.997009277344
25 813.341979980469
30 813.554016113281
35 826.557983398438
40 830.913024902344
45 836.443969726562
50 839.276977539062
55 841.632995605469
60 847.961975097656
65 855.481994628906
70 863.747009277344
75 862.987976074219
80 868.393981933594
85 877.486999511719
90 880.856994628906
95 883.607971191406
100 888.117980957031
};
\addlegendentry{CoFIM}
\addplot [thick, color0, mark=*, mark size=.5, mark options={solid}]
table {%
0 0
1 764.408996582031
5 869.810974121094
10 935.357971191406
15 980.879028320312
20 1019.35998535156
25 1043.63000488281
30 1064.22998046875
35 1082.26000976562
40 1101.78002929688
45 1116.46997070312
50 1131.47998046875
55 1141.72998046875
60 1154.66003417969
65 1165.34997558594
70 1179.31994628906
75 1189.35998535156
80 1200.07995605469
85 1208.77001953125
90 1216.75
95 1225.96997070312
100 1235.22998046875
};
\addlegendentry{CELF++}
\addplot [thick, color1, mark=*, mark size=.5, mark options={solid}]
table {%
0 0
1 764.408996582031
5 812.156005859375
10 859.956970214844
15 909.031982421875
20 944.596008300781
25 962.869995117188
30 1002.30999755859
35 1027.20300292969
40 1044.458984375
45 1067.82897949219
50 1087.5849609375
55 1096.89196777344
60 1111.71594238281
65 1125.57800292969
70 1145.87902832031
75 1157.54296875
80 1176.27294921875
85 1185.91198730469
90 1203.73706054688
95 1216.5400390625
100 1228.06604003906
};
\addlegendentry{Community-IM}
\end{axis}

\end{tikzpicture}}
  \subcaption{Wikipedia network}
\end{minipage}
\caption{Influence vs. budget $k$ for different networks under IC diffusion model and TV edge-weight model.}  \label{fig:ic-tv}
\end{figure*}
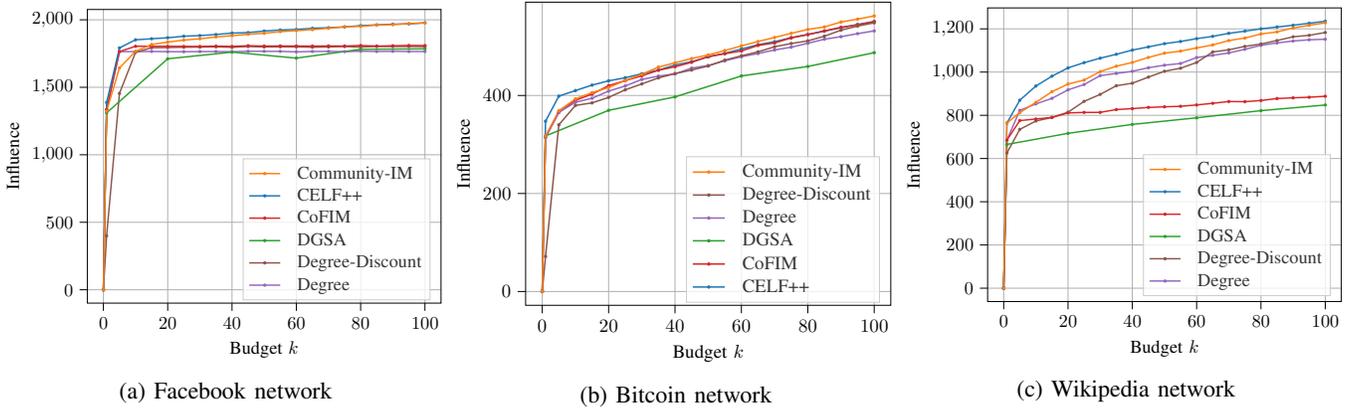

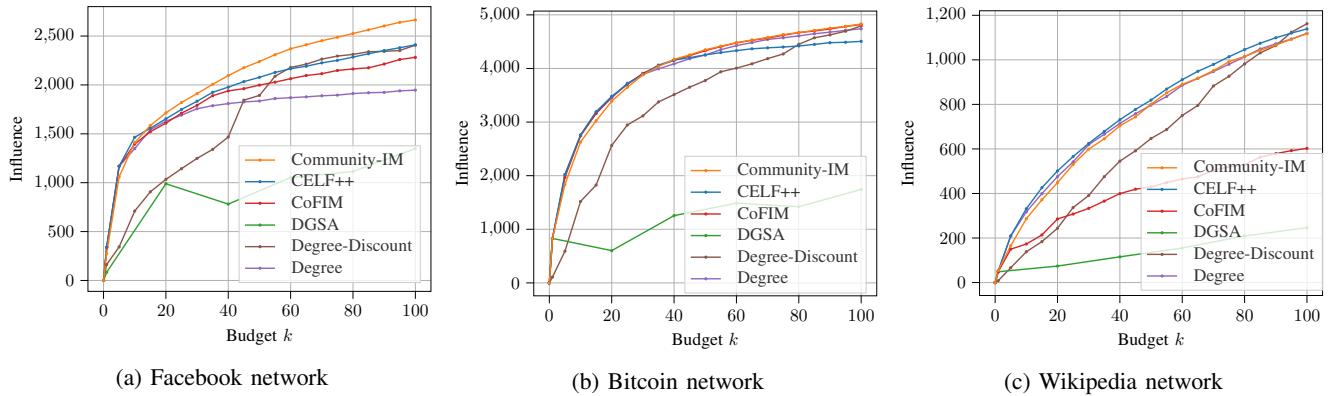
\begin{figure*}[t]
\centering
\begin{minipage}{.32\textwidth}
  \centering
  \resizebox{\textwidth}{!}{
\begin{tikzpicture}

\definecolor{color0}{rgb}{0.12156862745098,0.466666666666667,0.705882352941177}
\definecolor{color1}{rgb}{1,0.498039215686275,0.0549019607843137}
\definecolor{color2}{rgb}{0.172549019607843,0.627450980392157,0.172549019607843}
\definecolor{color3}{rgb}{0.83921568627451,0.152941176470588,0.156862745098039}
\definecolor{color4}{rgb}{0.580392156862745,0.403921568627451,0.741176470588235}
\definecolor{color5}{rgb}{0.549019607843137,0.337254901960784,0.294117647058824}

\begin{axis}[
legend cell align={left},
reverse legend,
legend style={
  fill opacity=0.8,
  draw opacity=1,
  text opacity=1,
  at={(0.97,0.03)},
  anchor=south east,
  draw=white!80!black
},
tick align=outside,
tick pos=left,
x grid style={white!69.0196078431373!black},
xlabel={Budget \(\displaystyle k\)},
xmajorgrids,
xmin=-5, xmax=105,
xtick style={color=black},
y grid style={white!69.0196078431373!black},
ylabel={Influence},
ymajorgrids,
ymin=-111.5672, ymax=2800.9112,
ytick style={color=black}
]

\addplot [thick, color4, mark=*, mark size=.5, mark options={solid}]
table {%
0 0
1 339.058013916016
5 1168.66003417969
10 1348.71398925781
15 1540.916015625
20 1628.11096191406
25 1690.92199707031
30 1756.76098632812
35 1787.32104492188
40 1808.89501953125
45 1826.07202148438
50 1835.35205078125
55 1860.60095214844
60 1869.5419921875
65 1878.06201171875
70 1889.09399414062
75 1895.57604980469
80 1911.93701171875
85 1920.2509765625
90 1923.78503417969
95 1939.18798828125
100 1946.59204101562
};
\addlegendentry{Degree}
\addplot [thick, color5, mark=*, mark size=.5, mark options={solid}]
table {%
0 0
1 162.029998779297
5 342.912994384766
10 709.833984375
15 906.64501953125
20 1034.15405273438
25 1143.498046875
30 1248.90795898438
35 1341.11401367188
40 1467.96398925781
45 1844.20703125
50 1892.74096679688
55 2088.6640625
60 2178.89404296875
65 2212.5
70 2263.9609375
75 2295.00708007812
80 2312.44799804688
85 2339.57592773438
90 2343.2060546875
95 2352.12890625
100 2406.45288085938
};
\addlegendentry{Degree-Discount}
\addplot [thick, color2, mark=*, mark size=.5, mark options={solid}]
table {%
0 0
1 83.6050033569336
20 989.388977050781
40 781.247985839844
60 1052.53698730469
80 1114.8330078125
100 1348.51293945312
};
\addlegendentry{DGSA}
\addplot [thick, color3, mark=*, mark size=.5, mark options={solid}]
table {%
0 0
1 339.058013916016
5 1168.66003417969
10 1395.34399414062
15 1520.88305664062
20 1608.86901855469
25 1711.40600585938
30 1791.75903320312
35 1889.46801757812
40 1939.14501953125
45 1961.67895507812
50 1998.41394042969
55 2028.46496582031
60 2063.662109375
65 2096.25805664062
70 2115.01904296875
75 2147.96606445312
80 2163.26489257812
85 2175.43994140625
90 2213.70092773438
95 2259.94506835938
100 2281.53002929688
};
\addlegendentry{CoFIM}
\addplot [thick, color0, mark=*, mark size=.5, mark options={solid}]
table {%
0 0
1 339.058013916016
5 1168.66003417969
10 1464.5
15 1557.18005371094
20 1652.9599609375
25 1749.38000488281
30 1833.43005371094
35 1924.59997558594
40 1977.43994140625
45 2033.67004394531
50 2077.8798828125
55 2127.68994140625
60 2166.13989257812
65 2191.0400390625
70 2225.72998046875
75 2251.3701171875
80 2284.6298828125
85 2319.39990234375
90 2351.78002929688
95 2380.36010742188
100 2410.28002929688
};
\addlegendentry{CELF++}
\addplot [thick, color1, mark=*, mark size=.5, mark options={solid}]
table {%
0 0
1 274.795013427734
5 1060.22302246094
10 1411.59497070312
15 1584.31103515625
20 1713.92895507812
25 1820.28894042969
30 1910.86206054688
35 2006.38696289062
40 2095.39892578125
45 2175.81103515625
50 2239.4599609375
55 2308.76391601562
60 2369.84692382812
65 2410.73388671875
70 2452.34399414062
75 2487.71801757812
80 2526.53100585938
85 2563.3779296875
90 2602.91088867188
95 2640.18896484375
100 2664.60205078125
};
\addlegendentry{Community-IM}
\end{axis}

\end{tikzpicture}}
  \subcaption{Facebook network}
\end{minipage}
\begin{minipage}{.32\textwidth}
  \centering
  \resizebox{\textwidth}{!}{
\begin{tikzpicture}

\definecolor{color0}{rgb}{0.12156862745098,0.466666666666667,0.705882352941177}
\definecolor{color1}{rgb}{1,0.498039215686275,0.0549019607843137}
\definecolor{color2}{rgb}{0.172549019607843,0.627450980392157,0.172549019607843}
\definecolor{color3}{rgb}{0.83921568627451,0.152941176470588,0.156862745098039}
\definecolor{color4}{rgb}{0.580392156862745,0.403921568627451,0.741176470588235}
\definecolor{color5}{rgb}{0.549019607843137,0.337254901960784,0.294117647058824}

\begin{axis}[
legend cell align={left},
reverse legend,
legend style={
  fill opacity=0.8,
  draw opacity=1,
  text opacity=1,
  at={(0.97,0.25)},
  anchor=east,
  draw=white!80!black
},
tick align=outside,
tick pos=left,
x grid style={white!69.0196078431373!black},
xlabel={Budget \(\displaystyle k\)},
xmajorgrids,
xmin=-5, xmax=105,
xtick style={color=black},
y grid style={white!69.0196078431373!black},
ylabel={Influence},
ymajorgrids,
ymin=-241.4653, ymax=5070.7713,
ytick style={color=black}
]

\addplot [thick, color4, mark=*, mark size=.5, mark options={solid}]
table {%
0 0
1 829.786010742188
5 2019.78002929688
10 2748.0009765625
15 3162.22705078125
20 3451.65698242188
25 3724.14306640625
30 3889.31909179688
35 3993.86303710938
40 4085.45092773438
45 4181.625
50 4250.39697265625
55 4351.44384765625
60 4426.21923828125
65 4482.81298828125
70 4541.2548828125
75 4572.14990234375
80 4608.423828125
85 4647.84716796875
90 4677.56005859375
95 4710.7001953125
100 4740.4609375
};
\addlegendentry{Degree}
\addplot [thick, color5, mark=*, mark size=.5, mark options={solid}]
table {%
0 0
1 107.68399810791
5 590.851013183594
10 1518.24096679688
15 1825.40295410156
20 2563.09692382812
25 2944.501953125
30 3116.05688476562
35 3375.63696289062
40 3512.56103515625
45 3650.34008789062
50 3774.544921875
55 3938.64501953125
60 4006.34204101562
65 4086.98193359375
70 4183.916015625
75 4270.60595703125
80 4451.10693359375
85 4572.18310546875
90 4626.35986328125
95 4695.19091796875
100 4793.51611328125
};
\addlegendentry{Degree-Discount}
\addplot [thick, color2, mark=*, mark size=.5, mark options={solid}]
table {%
0 0
1 829.786010742188
20 601.924987792969
40 1254.623046875
60 1488.15100097656
80 1420.84997558594
100 1742.90002441406
};
\addlegendentry{DGSA}
\addplot [thick, color3, mark=*, mark size=.5, mark options={solid}]
table {%
0 0
1 829.786010742188
5 1969.14904785156
10 2745.42895507812
15 3161.9970703125
20 3470.98608398438
25 3703.40600585938
30 3908.69091796875
35 4050.63208007812
40 4152.26708984375
45 4240.0859375
50 4332.0322265625
55 4408.48291015625
60 4477.3427734375
65 4525.794921875
70 4566.3759765625
75 4619.6728515625
80 4669.31982421875
85 4699.01806640625
90 4737.68701171875
95 4783.01318359375
100 4821.52783203125
};
\addlegendentry{CoFIM}
\addplot [thick, color0, mark=*, mark size=.5, mark options={solid}]
table {%
0 0
1 829.786010742188
5 2019.78002929688
10 2761.89990234375
15 3193.77001953125
20 3478.65991210938
25 3708.9599609375
30 3890.11010742188
35 4062.75
40 4156.22998046875
45 4201.7099609375
50 4253.72021484375
55 4296.6298828125
60 4333.2900390625
65 4367.18017578125
70 4385.16015625
75 4400.14013671875
80 4419.64990234375
85 4449.25
90 4480.93017578125
95 4489.2099609375
100 4505.64990234375
};
\addlegendentry{CELF++}
\addplot [thick, color1, mark=*, mark size=.5, mark options={solid}]
table {%
0 0
1 829.786010742188
5 1836.08996582031
10 2627.30590820312
15 3023.8779296875
20 3388.71801757812
25 3647.7490234375
30 3881.12109375
35 4038.748046875
40 4171.05322265625
45 4253.4228515625
50 4348.3310546875
55 4416.28076171875
60 4481.7529296875
65 4531.27099609375
70 4579.15478515625
75 4632.73779296875
80 4668.89794921875
85 4711.39208984375
90 4751.73779296875
95 4786.966796875
100 4829.30615234375
};
\addlegendentry{Community-IM}
\end{axis}

\end{tikzpicture}}
  \subcaption{Bitcoin network}
\end{minipage}
\begin{minipage}{.32\textwidth}
  \centering
  \resizebox{\textwidth}{!}{
\begin{tikzpicture}

\definecolor{color0}{rgb}{0.12156862745098,0.466666666666667,0.705882352941177}
\definecolor{color1}{rgb}{1,0.498039215686275,0.0549019607843137}
\definecolor{color2}{rgb}{0.172549019607843,0.627450980392157,0.172549019607843}
\definecolor{color3}{rgb}{0.83921568627451,0.152941176470588,0.156862745098039}
\definecolor{color4}{rgb}{0.580392156862745,0.403921568627451,0.741176470588235}
\definecolor{color5}{rgb}{0.549019607843137,0.337254901960784,0.294117647058824}

\begin{axis}[
legend cell align={left},
reverse legend,
legend style={
  fill opacity=0.8,
  draw opacity=1,
  text opacity=1,
  at={(0.47,0.5)},
  anchor=north west,
  draw=white!80!black
},
tick align=outside,
tick pos=left,
x grid style={white!69.0196078431373!black},
xlabel={Budget \(\displaystyle k\)},
xmajorgrids,
xmin=-5, xmax=105,
xtick style={color=black},
y grid style={white!69.0196078431373!black},
ylabel={Influence},
ymajorgrids,
ymin=-58.1105, ymax=1220.3205,
ytick style={color=black}
]

\addplot [thick, color4, mark=*, mark size=.5, mark options={solid}]
table {%
0 0
1 48.4749984741211
5 209.194000244141
10 317.221008300781
15 399.341003417969
20 475.794006347656
25 541.783020019531
30 619.002990722656
35 665.958984375
40 714.192993164062
45 757.945007324219
50 796.697998046875
55 835.388977050781
60 885.638000488281
65 916.575012207031
70 947.210021972656
75 980.174987792969
80 1013.59698486328
85 1048.76599121094
90 1071.80395507812
95 1093.4189453125
100 1118.68505859375
};
\addlegendentry{Degree}
\addplot [thick, color5, mark=*, mark size=.5, mark options={solid}]
table {%
0 0
1 8.71000003814697
5 66.4199981689453
10 138.257995605469
15 183.835006713867
20 243.84700012207
25 336.730010986328
30 390.769012451172
35 475.299987792969
40 545.213989257812
45 591.541015625
50 646.898986816406
55 687.195007324219
60 749.953979492188
65 795.418029785156
70 882.669982910156
75 926.382019042969
80 981.745971679688
85 1031.26403808594
90 1063.86694335938
95 1123.92700195312
100 1162.2099609375
};
\addlegendentry{Degree-Discount}
\addplot [thick, color2, mark=*, mark size=.5, mark options={solid}]
table {%
0 0
1 48.4749984741211
20 74.0670013427734
40 115.47200012207
60 155.070007324219
80 209.311004638672
100 246.263000488281
};
\addlegendentry{DGSA}
\addplot [thick, color3, mark=*, mark size=.5, mark options={solid}]
table {%
0 0
1 48.4749984741211
5 149.901992797852
10 173.029006958008
15 213.856994628906
20 285.5419921875
25 307.688995361328
30 333.221008300781
35 365.739013671875
40 399.522003173828
45 419.161010742188
50 429.592010498047
55 449.049011230469
60 464.894012451172
65 474.971984863281
70 503.820007324219
75 519.742004394531
80 528.612976074219
85 563.185974121094
90 578.953979492188
95 592.247009277344
100 602.331970214844
};
\addlegendentry{CoFIM}
\addplot [thick, color0, mark=*, mark size=.5, mark options={solid}]
table {%
0 0
1 53.6209983825684
5 209.585006713867
10 331.610992431641
15 426.169006347656
20 501.222991943359
25 565.523986816406
30 624.567016601562
35 678.567993164062
40 731.075988769531
45 777.335021972656
50 819.258972167969
55 868.919006347656
60 910.940979003906
65 948.523010253906
70 979.361999511719
75 1013.88000488281
80 1046.14001464844
85 1074.35998535156
90 1099.06005859375
95 1120.18994140625
100 1138.64001464844
};
\addlegendentry{CELF++}
\addplot [thick, color1, mark=*, mark size=.5, mark options={solid}]
table {%
0 0
1 53.6209983825684
5 165.149993896484
10 286.497985839844
15 371.598999023438
20 448.899993896484
25 529.940002441406
30 598.843994140625
35 645.543029785156
40 703.523010253906
45 743.56201171875
50 800.831970214844
55 851.89501953125
60 890.151000976562
65 917.601989746094
70 951.528991699219
75 991.392028808594
80 1015.40600585938
85 1042.65698242188
90 1066.13598632812
95 1092.53601074219
100 1117.35705566406
};
\addlegendentry{Community-IM}
\end{axis}

\end{tikzpicture}}
  \subcaption{Wikipedia network}
\end{minipage}
\caption{Influence vs. budget $k$ for different networks under LT diffusion model and WC edge-weight model.} \label{fig:lt-wc}
\end{figure*}

For different networks under different diffusion models and edge-weight models,   
\begin{itemize}
    \item Figures~\ref{fig:ic-wc}-\ref{fig:lt-wc} show the influences of chosen seed sets using different algorithms for different values of budget $k$.
    Figure~\ref{fig:ic-wc} shows the results for IC diffusion model and WC edge-weight model, Figure~\ref{fig:ic-tv} shows the results for IC diffusion model and TV edge-weight model, and Figure~\ref{fig:lt-wc} shows the results for LT diffusion model and WC edge-weight model.
    \item Table~\ref{tab:max_exp_influences} and Table~\ref{tab:run_times} show the influences and run-times, respectively for budget $k = 100$ for different algorithms. 
\end{itemize}

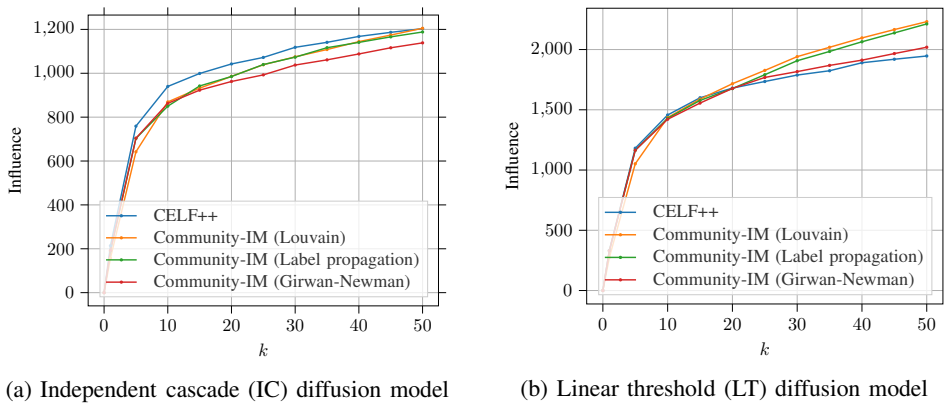
\begin{figure*}[h]
\centering
\begin{minipage}{.33\textwidth}
  \centering
  \resizebox{\textwidth}{!}{
\begin{tikzpicture}

\definecolor{color0}{rgb}{0.12156862745098,0.466666666666667,0.705882352941177}
\definecolor{color1}{rgb}{1,0.498039215686275,0.0549019607843137}
\definecolor{color2}{rgb}{0.172549019607843,0.627450980392157,0.172549019607843}
\definecolor{color3}{rgb}{0.83921568627451,0.152941176470588,0.156862745098039}

\begin{axis}[
legend cell align={left},
legend style={fill opacity=0.8, draw opacity=1, text opacity=1, at={(0.97,0.03)}, anchor=south east, draw=white!80!black},
tick align=outside,
tick pos=both,
x grid style={white!69.0196078431373!black},
xlabel={\(\displaystyle k\)},
xmajorgrids,
xmin=-2.5, xmax=52.5,
xtick style={color=black},
y grid style={white!69.0196078431373!black},
ylabel={Influence},
ymajorgrids,
ymin=-60.25625, ymax=1265.38125,
ytick style={color=black}
]
\addplot [thick, color0, mark=*, mark size=.5, mark options={solid}]
table {%
0 0
1 214.046005249023
5 758.893981933594
10 939.913024902344
15 999.086975097656
20 1042.43005371094
25 1072.15002441406
30 1118.31005859375
35 1140.96997070312
40 1167.93994140625
45 1186.55004882812
50 1203.42004394531
};
\addlegendentry{CELF++}
\addplot [thick, color1, mark=*, mark size=.5, mark options={solid}]
table {%
0 0
1 157.576995849609
5 642.02001953125
10 869.364013671875
15 932.114990234375
20 985.815979003906
25 1038.90795898438
30 1074.95202636719
35 1108.251953125
40 1144.28405761719
45 1174.126953125
50 1205.125
};
\addlegendentry{Community-IM (Louvain)}
\addplot [thick, color2, mark=*, mark size=.5, mark options={solid}]
table {%
0 0
1 189.774993896484
5 703.010986328125
10 850.031005859375
15 942.067993164062
20 985.236999511719
25 1039.80102539062
30 1073.38793945312
35 1116.51696777344
40 1140.75903320312
45 1165.88793945312
50 1188.15698242188
};
\addlegendentry{Community-IM (Label propagation)}
\addplot [thick, color3, mark=*, mark size=.5, mark options={solid}]
table {%
0 0
1 192.02799987793
5 703.976989746094
10 862.900024414062
15 923.046997070312
20 962.547973632812
25 992.611022949219
30 1037.48596191406
35 1061.17602539062
40 1088.04699707031
45 1116.17797851562
50 1138.59094238281
};
\addlegendentry{Community-IM (Girwan-Newman)}
\end{axis}
\end{tikzpicture}}
  \subcaption{Independent cascade (IC) diffusion model}
\end{minipage}\hspace{.5cm}
\begin{minipage}{.33\textwidth}
  \centering
  \resizebox{\textwidth}{!}{
\begin{tikzpicture}

\definecolor{color0}{rgb}{0.12156862745098,0.466666666666667,0.705882352941177}
\definecolor{color1}{rgb}{1,0.498039215686275,0.0549019607843137}
\definecolor{color2}{rgb}{0.172549019607843,0.627450980392157,0.172549019607843}
\definecolor{color3}{rgb}{0.83921568627451,0.152941176470588,0.156862745098039}

\begin{axis}[
legend cell align={left},
legend style={
  fill opacity=0.8,
  draw opacity=1,
  text opacity=1,
  at={(0.97,0.03)},
  anchor=south east,
  draw=white!80!black
},
tick align=outside,
tick pos=left,
x grid style={white!69.0196078431373!black},
xlabel={\(\displaystyle k\)},
xmajorgrids,
xmin=-2.5, xmax=52.5,
xtick style={color=black},
y grid style={white!69.0196078431373!black},
ylabel={Influence},
ymajorgrids,
ymin=-111.5672, ymax=2342.9112,
ytick style={color=black}
]
\addplot [thick, color0, mark=*, mark size=.5, mark options={solid}]
table {%
0 0
1 332.5830078125
5 1180.93994140625
10 1456.71997070312
15 1599.66003417969
20 1679.9599609375
25 1734.81005859375
30 1788.18994140625
35 1824.02001953125
40 1890.38000488281
45 1919.43994140625
50 1946.41003417969
};
\addlegendentry{CELF++}
\addplot [thick, color1, mark=*, mark size=.5, mark options={solid}]
table {%
0 0
1 274.795013427734
5 1051.94799804688
10 1435.09802246094
15 1588.47302246094
20 1716.34704589844
25 1826.81298828125
30 1941.05200195312
35 2018.78002929688
40 2096.39306640625
45 2165.74096679688
50 2231.34399414062
};
\addlegendentry{Community-IM (Louvain)}
\addplot [thick, color2, mark=*, mark size=.5, mark options={solid}]
table {%
0 0
1 329.915985107422
5 1164.07800292969
10 1429.26403808594
15 1577.66394042969
20 1676.15100097656
25 1791.13598632812
30 1907.16198730469
35 1984.1669921875
40 2064.50708007812
45 2138.087890625
50 2212.56591796875
};
\addlegendentry{Community-IM (Label propagation)}
\addplot [thick, color3, mark=*, mark size=.5, mark options={solid}]
table {%
0 0
1 329.915985107422
5 1164.07800292969
10 1421.36401367188
15 1555.76000976562
20 1677.06005859375
25 1768.52600097656
30 1816.89294433594
35 1868.44396972656
40 1911.96704101562
45 1966.501953125
50 2019.08203125
};
\addlegendentry{Community-IM (Girwan-Newman)}
\end{axis}

\end{tikzpicture}}
  \subcaption{Linear threshold (LT) diffusion model}
\end{minipage}
\caption{Influence vs. budget $k$ for the Facebook network under different diffusion models and WC edge-weight model.} \label{fig:facebook-com}
\end{figure*}

For the Facebook network under different diffusion models, and WC edge-weight model for different community detection methods as \texttt{com-method} in Community-Detection step of the proposed framework,
\begin{itemize}
    \item Figure~\ref{fig:facebook-com} shows the influences of chosen seed sets using different algorithms for different values of $k$.
    \item Table~\ref{tab:max_exp_influences_com} shows the modularity scores, the number of communities, and the influences and run-times for budget $k=50$ for \text{Community-IM} and \text{CELF++}.
\end{itemize}

\begin{table*}[t] 
    \centering
	\caption{Comparison of influences for budget $k=100$. } \label{tab:max_exp_influences}
	\begin{tabular}{p{1.6cm} p{1.7cm} p{1.2cm} r r r r r r}
		\hline \hline
		Diffusion model & Edge-weight model & Network & \text{Community-IM} & \text{CELF++} & \text{CoFIM} & DSGA & Degree & Degree-Discount\\
		\hline
        \multirow{7}{4em}{Independent cascade} & \multirow{4}{4em}{Weighted cascade} & Facebook & 1,378 & 1,406 & 1,237 & 846 & 1,092 & 1,289 \\
                                              & & Bitcoin & 3,693 & 3,493 & 3,679 & 1,643 & 3,596 & 3,625\\ 
                                              & & Wikipedia & 873 & 877 & 528 & 213 & 866 & 878\\ 
                                              & & Epinions & 14,706 & 14,043 & 12,315 & 2,439 & 13,458 & 13,771\\ \cline{2-9} 
		                                      & \multirow{4}{*}{Trivalency} & Facebook & 1,977 & 1,977 & 1,809 & 1,305 & 1,765 & 1,801 \\ 
                                              & & Bitcoin & 562 & 551 & 551 & 487 & 532 & 548\\
                                              & & Wikipedia & 1,228 & 1,235 & 888 & 848 & 1,152 & 1,183\\ 
                                            \cline{2-9} 
		\hline
		\multirow{3}{4em}{Linear threshold} & \multirow{3}{4em}{Weighted cascade} & Facebook & 2,231 & 1,946 & 1,936 & 969 & 1,835 & 2,000\\
                                              & & Bitcoin & 4,829 & 4,506 & 4,822 & 1,743 & 4,740 & 4,794 \\
                                              & & Wikipedia & 1,117 & 1,139 & 602 & 246 & 1,119 & 1,162 \\ 
                                            \cline{2-9}
	    \hline
	    \hline
	\end{tabular}
\end{table*}

\begin{table*}[t] 
    \centering
	\caption{Comparison of run-times (in seconds) for budget $k=100$.} \label{tab:run_times}
	\begin{tabular}{p{1.6cm} p{1.7cm} p{1.2cm} r r r r}
		\hline \hline
		Diffusion model & Edge-weight model & Network &   \text{Community-IM} &  \text{CELF++} & CoFIM & DSGA \\
		\hline
        \multirow{7}{4em}{Independent cascade} & \multirow{4}{4em}{Weighted cascade} & Facebook & 3,782 & 17,359 & 547 & 9,1267\\
                                              & & Bitcoin & 850 & 10,859 & 35 & 20,825\\ 
                                              & & Wikipedia & 3,477 & 2,660 & 213 & 18,447\\ 
                                              & & Epinions & 16,465 & 250,241 & 7,397 & 267,796\\ 
                                              \cline{2-7}
		                                       & \multirow{3}{*}{Trivalency} & Facebook & 7,195  & 74,684 & 567 & 312,948 \\ 
                                              & & Bitcoin & 576 & 7,818 & 35 & 34,453\\
                                            \cline{2-7}
		\hline
		\multirow{3}{4em}{Linear threshold} & \multirow{3}{4em}{Weighted cascade} & Facebook & 8,545 & 46,771 & 554 & 65,391\\
                                               & & Bitcoin & 1,077 & 45,747 & 36 & 62,184 \\
                                              & & Wikipedia & 4,628 & 5,940 & 224 & 23,307\\ 
                                            \cline{2-7}
	    \hline
	    \hline
	\end{tabular}
\end{table*}

\begin{table*}[h] 
    \centering
    \caption{Comparison of influences and run-times (in seconds) for budget $k=50$ for the Facebook network under WC edge-weight model for different community detection methods.} \label{tab:max_exp_influences_com}
    \begin{tabular}{p{1.4cm} p{2.1cm} r r r r r r}
    \hline \hline
        & & & & \multicolumn{2}{c}{Influence} & \multicolumn{2}{c}{Run-times (in seconds)}\\
        \hline
    Diffusion model & Community detection method & No. of communities & Modularity score  & \text{Community-IM} & \text{CELF++} & \text{Community-IM} & \text{CELF++}\\
    \hline
    \multirow{3}{4em}{Independent cascade} & Louvain & 18 & 0.8678  & 1,205 & 1,203 & 3,069 & 14,077\\
                                           & Label propagation & 11 & 0.7368  & 1,188 & 1,203 & 4,068 & 14,077\\
                                           & Girvan-Newman & 2 & 0.0439 & 1,139 & 1,203 & 14,221 & 14,077\\ \cline{2-6}
    \hline
    \multirow{3}{4em}{Linear threshold} & Louvain & 18 & 0.8304 & 2,231 & 1,946 & 7,224 & 38,968\\
                                          & Label propagation & 11 & 0.7368 & 2,213 & 1,946 & 12,961 & 38,968\\
                                          & Girvan-Newman & 2 & 0.0439 & 2,019 & 1,946 & 34,606 & 38,968\\ \cline{2-6}
    \hline
    \hline
    \end{tabular}
\end{table*}

\subsection{Discussion}
\subsubsection{Overview}  
The proposed framework (Community-IM) achieves either marginally lower, equal, or higher influence compared to CELF++, and  achieves better influence compared to all other algorithms. This performance in terms of influence improves  as the budget increases. The proposed framework brings savings in terms of run-time as compared to the simulation-based methods.  The community structures with higher modularity lead the proposed framework to perform better in terms of run-time and influence. Moreover, these observations vary across different networks, diffusion models, edge-weight models, and budgets.

\subsubsection{Performance in terms of influence} For low budgets, the influence for \text{Community-IM} (orange) is marginally lower than that for \text{CELF++} (blue). However, for high budgets, the influence for \text{Community-IM} is the same or higher than that for \text{CELF++}. Furthermore, the influence for \text{Community-IM} is higher compared to the rest of the algorithms. We observe this trend for the Facebook, Bitcoin, and Epinions networks under different diffusion models and different edge-weight models from Figures~\ref{fig:ic-wc}(a), \ref{fig:ic-wc}(b), and \ref{fig:ic-wc}(d), Figures~\ref{fig:ic-tv}(a) and \ref{fig:ic-tv}(b), and Figures~\ref{fig:lt-wc}(a) and \ref{fig:lt-wc}(b). 

For all budgets, the influence for \text{Community-IM} (orange) is marginally lower than that for \text{CELF++} (blue). However, the gap between the influence for \text{Community-IM} and that for \text{CELF++} decreases as the budget $k$ increases. Furthermore, the influence for \text{Community-IM} is higher compared to the rest of the algorithms. We observe this trend for the Wikipedia network under different diffusion and edge-weight models from Figures~\ref{fig:ic-wc}(c), \ref{fig:ic-tv}(c), and \ref{fig:lt-wc}(c). 

Note that by design, the proposed community-aware framework gives preference to the community-level influential nodes while building its nested solution using progressive budgeting (Algorithm~\ref{alg:pb}). However, when the budget is large, depending on their community-level influence, the network-level influential nodes are also selected. On the contrary, CELF++ prefers network-level influential nodes while building its nested solution. Hence, the proposed framework takes advantage of the community-level influence ordering of nodes early on. However,  network-level celebrities may not be equally popular within each community. Hence, particularly for low budgets, the proposed framework selects only the community-level influential nodes. However, when the budget is large, it starts to pick the network-level influential nodes as well. This explains why the performance of the proposed algorithm in terms of influence gets better as the budget increases. Such a trend gets more pronounced for networks that have some extremely (network-level) influential nodes (e.g. the Facebook and Epinions networks) that are not selected initially for small values of the budget but included later for high budgets. 

Moreover, Table~\ref{tab:max_exp_influences} shows that for each network, the influence of the chosen seed set of size 100 using \text{Community-IM} is close to or even better than the same for \text{CELF++} under different diffusion models and edge-weight models. 

\subsubsection{Performance in terms of run-time} Table~\ref{tab:run_times} shows that the proposed framework brings savings in terms of run-time as compared to the simulation-based methods (\text{CELF++}, and \textsc{DSGA}) across different networks, diffusion models, and edge-weight models. Moreover, these run-time savings are more pronounced for larger networks. The gains in terms of run-time also vary across diffusion models and edge-weight models. We observe the highest gains for IC diffusion model with TV edge-weight model and the least gains for the IC diffusion model with WC edge-weight model.

\subsubsection{Effect of the community structure on the performance of the proposed framework} Based on Figure~\ref{fig:facebook-com}, we observe that the community structures with higher values of modularity (obtained using the Louvain and Label Propagation methods) lead the proposed framework to do better in terms of influence as compared to the community structures with lower values of modularity (obtained using the Girvan-Newman method \cite{girvan2002community}). Furthermore, for all budgets, the influences for \text{Community-IM} with Louvain method and \text{Community-IM} with Label propagation method are close to each other which can be attributed to the fact that the modularity scores of the partitions obtained by these two methods are quite close.

Table~\ref{tab:max_exp_influences_com} shows that the influence for the budget of $k=50$, using \text{Community-IM} is close to or even better than the same for \text{CELF++} for different choices of community detection methods under different diffusion models and WC edge-weight model. Furthermore, the performance of \text{Community-IM} compared to \text{CELF++} in terms of influence and run-time improves as the modularity of the partition and the number of communities increase. Note that, for the proposed framework, the Louvain method is the best choice of community detection method while the Girvan-Newman method performs the worst. The Louvain method partitions the graph into 18 communities with the largest community having 523 nodes (approximately 10\% of the size of the entire network). Hence, \text{Community-IM} does not come across any giant component (causing lengthier diffusions) while estimating the within-community influence. Contrary to this, the Girvan-Newman method partitions the network into just two communities with the largest community having 3,833 nodes (very close to the size of the entire network). This makes the within-community diffusions take longer to finish while using the communities obtained using the Girvan-Newman method. This explains why Community-IM with the Girvan-Newman method runs slower as compared to the same with the Louvain method.
\section{Conclusion and Future Work} \label{section5}
For solving the problem of influence maximization on social networks, we leveraged the inherent community structure of a network and proposed a novel community-aware framework for maximizing the spread of influence through a social network in a fast manner. Based on our experiments, we conclude that the proposed framework outperforms the standard simulation-based methods in terms of run-time and the heuristic methods in terms of influence. As the proposed method leverages the inherent community structure of the network, we also studied the effect of the community structure on the performance of our framework. Based on our experiments, we conclude that the community structures with higher modularity lead the proposed framework to perform better in terms of run-time and influence. Among the methods considered in this paper, we find the Louvain method \cite{blondel2008fast} works best for our framework.

We point out two limitations of our method. First, our method requires the communities learned during Step 1 to be non-overlapping. However, in general, a social network may have overlapping communities. Second, our method does not explicitly account for the inter-community influence while generating the candidate solutions during Step 2. In the future, we want to extend our method to handle overlapping community structures and explicitly account for the inter-community influence. Other future directions are to extend the proposed community-aware framework to competitive influence maximization \cite{bharathi2007competitive}, data-based influence maximization \cite{goyal2011data}, and full-bandit online influence maximization \cite{agarwal2022stochastic,nie2022explore}.
\bibliographystyle{IEEEtran}
\bibliography{refs.bib}

\vfill

\vspace{-5mm}

\begin{IEEEbiographynophoto}{Abhishek K. Umrawal} is currently a Ph.D. Candidate in the School of Industrial Engineering at Purdue University, West Lafayette, IN 47907, USA, and a Visiting Lecturer in the Department of Computer Science and Electrical Engineering at the University of Maryland, Baltimore County, MD 21250, USA. He received an M.Sc. degree in Statistics from the Indian Institute of Technology Kanpur, India, and an M.S. degree in Economics from Purdue University. His research interests include causality, reinforcement learning, optimization, and network science with applications to social networks and intelligent transportation.
\end{IEEEbiographynophoto}

\vspace{-5mm}

\begin{IEEEbiographynophoto}{Christopher J. Quinn} (Member, IEEE) is currently an Assistant Professor in the Department of Computer Science at Iowa State University, Ames, IA 50011, USA. He received a B.S. degree in Engineering Physics from Cornell University, and M.S. and Ph.D. degrees in Electrical and Computer Engineering from the University of Illinois at Urbana-Champaign. His current research interests include machine learning, information theory, and network science, with applications to neuroscience and social networks.
\end{IEEEbiographynophoto}

\vspace{-5mm}

\begin{IEEEbiographynophoto}{Vaneet Aggarwal} (Senior Member, IEEE) is currently a Full Professor at  Purdue University, West Lafayette, IN 47907, USA. He received a B.Tech. degree from the Indian Institute of Technology Kanpur, India, and M.A. and Ph.D. degrees from Princeton University, all in Electrical Engineering.  His current research interests include machine learning and its applications in networking, transportation, and quantum systems.
\end{IEEEbiographynophoto}

\clearpage
\clearpage
\onecolumn
\appendices
\setcounter{page}{1}
\section{Table of Notations} \label{sec:notations}
\begin{table}[ht] 
    \centering
	\caption{Table of notations.} \label{tab:notations}
	\begin{tabular}{l l}
		\hline \hline
		Symbol & Explanation\\
		\hline
		$\Omega$ & Ground set. \\
		$2^{\Omega}$ & Set of all subsets of $\Omega$.\\
		$G = (V, E)$ & Directed graph. \\
		$V = (v_1,\dots ,v_n)$ & Set of vertices or nodes.\\
		$n$ & Size of $V$.\\
		$E = (e_1,\dots ,e_n)$ & Set of directed edges where $e_i,i=1,\dots,n$ is are ordered pairs of nodes.\\
		$p_{v,w}$ & Weight of the edge $v \rightarrow w$.\\
		$\partial v$ & Set of neighbors of node $v$.\\
		$Y_t^{(v)}$ & Activation/state of node $v$ at time $t$.\\
		$k$ & Budget.\\
		$\sigma(S)$ & Influence of a set $S$ of nodes.\\
		$c$ & Number of communities.\\
		\texttt{com-method} & Community detection method.\\
		\texttt{sol-method} & Influence maximization method.\\
		$\{G_1,\dots G_c\}$ & A partition of $G$ with $c$ sub-graphs that are $G_1, \dots G_c$.\\
		$\{V_1,\dots V_c\}$ & Set of sets of vertices for all sub-graphs in the partition $\{G_1,\dots G_c\}$.\\
		$n_i$ & Size of $V_i,i=1,\dots,c$.\\
		$\{E_1,\dots E_c\}$ & Set of sets of edges for all sub-graphs in the partition $\{G_1,\dots G_c\}$.\\
		$Q$ & Modularity of a network partition.\\
		$S_{i,j}$ & Best seed set of size $j$ ($j=1,\dots,k$) from community $i$ ($i=1,\dots,c$).\\ 
		$\sigma_i(S_{i,j})$ & Influence of $S_{i,j}$ within community $i$ ($i=1,\dots,c$).\\
		$\mathcal{S}_i$ &  Set of all candidate solutions from community $i$ = $\{S_{i,j}: j=1,\dots,k\}$.\\
		$\Sigma_i$ &  Influences of all candidate solutions from community $i$ = $\{\sigma_i(S_{i,j}): j=1,\dots,k\}$.\\
		$\mathcal{S}$ & Set of sets of all candidate solutions from all communities = $\{\mathcal{S}_i: i=1,\dots,c\}$.\\ 
		$\Sigma$ & Set of sets of influences of all candidate solutions from all communities = $\{\Sigma_i: i=1,\dots,c\}$.\\
		$S^*$ & Final solution using the proposed framework.\\
	    \hline
	    \hline
	\end{tabular}
\end{table}

\section{An illustrative example of progressive budgeting} \label{subsec:pbexample}
In this section, we provide an illustrative example of progressive budgeting. After executing the Community-Detection and the Generate-Candidates steps of the proposed framework, we obtain the following output.
\begin{align*}
S_{i,j} &= \text{Candidate set of size } j \text{ from community } i, \\
\sigma_{i,j}:= \sigma_i(S_{i,j}) &= \text{Influence of } S_{i,j} \text{ within community } i,\\ 
i &= 1,\dotsc, j = 1,\dots,k.
\end{align*}
Let the budget, $k = 4$. No. of communities, $c = 5$. The influences of different candidate sets within different communities are given in Table~\ref{tab:pb-inputs}(a). For every $i=1,\dots,c;j=1,\dots,k$, we calculate the marginal influences as $m_{i,j} := \sigma_i(S_{i,j}) - \sigma_i(S_{i,j-1})$, where $\sigma_i(S_{i,0}) = 0, \forall i$. The marginal influences for the influences given in Table~\ref{tab:pb-inputs}(a) are provided in Table~\ref{tab:pb-inputs}(b).
\begin{table}[ht]
\centering
\begin{minipage}{.45\textwidth}
    \centering
    \begin{tabular}{c|c|c|c|c|c|}
        \multicolumn{6}{c}{Influence}\\ \cline{2-6}
        & $i$ & $\sigma_{i,1}$ & $\sigma_{i,2}$ & $\sigma_{i,3}$ & $\sigma_{i,4}$\\ \cline{2-6}
        \multirow{5}{*}{\rotatebox{90}{\small{Community}}} & 1 & 8 & 14 & 18 & 21\\ \cline{2-6}
        & 2 & 5 & 10 & 14 & 15\\ \cline{2-6}
        & 3 & 9 & 14 & 16 & 17\\ \cline{2-6}
        & 4 & 7 & 12 & 16 & 18\\ \cline{2-6}
        & 5 & 5 & 9 & 11 & 11\\ \cline{2-6}
    \end{tabular}
    \subcaption{Influences of candidate sets after step (ii).}
\end{minipage}
\begin{minipage}{.45\textwidth}
    \centering
    \begin{tabular}{c|c|c|c|c|c|}
        \multicolumn{6}{c}{Marginal Influence}\\ \cline{2-6}
        & $i$ & $m_{i,1}$ & $m_{i,2}$ & $m_{i,3}$ & $m_{i,4}$\\ \cline{2-6}
        \multirow{5}{*}{\rotatebox{90}{\small{Community}}} & 1 & 8 & 6 & 4 & 3\\ \cline{2-6}
        & 2 & 5 & 5 & 4 & 1\\ \cline{2-6}
        & 3 & 9 & 5 & 2 & 1\\ \cline{2-6}
        & 4 & 7 & 5 & 4 & 2\\ \cline{2-6}
        & 5 & 5 & 4 & 2 & 0\\ \cline{2-6}
    \end{tabular}
    \subcaption{Marginal Influences of candidate sets after step (ii).}
\end{minipage}
\caption{An example input for progressive budgeting.} \label{tab:pb-inputs}
\end{table}

The progressive budgeting scheme for the example in Table~\ref{tab:pb-inputs} is explained in Table~\ref{tab:pb-illustration}. At any iteration, the circled cells (\circled{$\cdot$}) are the ones whose maximum is to be obtained. An asterisk ($*$) is placed before the maximum value at the current iteration. The superscript(s) on any community label represents the nodes selected from that community (ordered based on the ordering in the corresponding candidate set obtained in step (ii)). 

\begin{table}[ht]
\centering
\begin{minipage}{.45\textwidth}
    \centering
    \begin{tabular}{c|l|r|r|r|r|}
        \multicolumn{6}{c}{Marginal Influence}\\ \cline{2-6}
        & $i$ & $m_{i,1}$ & $m_{i,2}$ & $m_{i,3}$ & $m_{i,4}$\\ \cline{2-6}
        \multirow{5}{*}{\rotatebox{90}{\small{Community}}} & 1 & \circled{8} & 6 & 4 & 3\\ \cline{2-6}
        & 2 & \circled{5} & 5 & 4 & 1\\ \cline{2-6}
        & $3^{1}$ & $*$ \circled{9} & 5 & 2 & 1\\ \cline{2-6}
        & 4 & \circled{7} & 5 & 4 & 2\\ \cline{2-6}
        & 5 & \circled{5} & 4 & 2 & 0\\ \cline{2-6}
    \end{tabular}
    \subcaption{Iteration 1: Allocating the first unit.}
\end{minipage} \vspace{.2cm}
\begin{minipage}{.45\textwidth}
    \centering
    \begin{tabular}{l|c|r|r|r|r|}
        \multicolumn{6}{c}{Marginal Influence}\\ \cline{2-6}
        & $i$ & $m_{i,1}$ & $m_{i,2}$ & $m_{i,3}$ & $m_{i,4}$\\ \cline{2-6}
        \multirow{5}{*}{\rotatebox{90}{\small{Community}}} & $1^{1}$ & $*$ \circled{8} & 6 & 4 & 3\\ \cline{2-6}
        & 2 & \circled{5} & 5 & 4 & 1\\ \cline{2-6}
        & $3^{1}$ & 9 & \circled{5} & 2 & 1\\ \cline{2-6}
        & 4 & \circled{7} & 5 & 4 & 2\\ \cline{2-6}
        & 5 & \circled{5} & 4 & 2 & 0\\ \cline{2-6}
    \end{tabular}
    \subcaption{Iteration 2: Allocating the second unit.}
\end{minipage} 
\begin{minipage}{.45\textwidth}
    \centering
    \begin{tabular}{l|c|r|r|r|r|}
        \multicolumn{6}{c}{Marginal Influence}\\ \cline{2-6}
        & $i$ & $m_{i,1}$ & $m_{i,2}$ & $m_{i,3}$ & $m_{i,4}$\\ \cline{2-6}
        \multirow{5}{*}{\rotatebox{90}{\small{Community}}} & $1^{1}$ & 8 & \circled{6} & 4 & 3\\ \cline{2-6}
        & 2 & \circled{5} & 5 & 4 & 1\\ \cline{2-6}
        & $3^{1}$ & 9 & \circled{5} & 2 & 1\\ \cline{2-6}
        & $4^{1}$ & $*$  \circled{7} & 5 & 4 & 2\\ \cline{2-6}
        & 5 & \circled{5} & 4 & 2 & 0\\ \cline{2-6}
    \end{tabular}
    \subcaption{Iteration 3: Allocating the third unit.}
\end{minipage} \vspace{.2cm}
\begin{minipage}{.45\textwidth}
    \centering
    \begin{tabular}{l|c|r|r|r|r|}
        \multicolumn{6}{c}{Marginal Influence}\\ \cline{2-6}
        & $i$ & $m_{i,1}$ & $m_{i,2}$ & $m_{i,3}$ & $m_{i,4}$\\ \cline{2-6}
        \multirow{5}{*}{\rotatebox{90}{\small{Community}}} & $1^{1,2}$ & 8 & $*$ \circled{6} & 4 & 3\\ \cline{2-6}
        & 2 & \circled{5} & 5 & 4 & 1\\ \cline{2-6}
        & $3^{1}$ & 9 & \circled{5} & 2 & 1\\ \cline{2-6}
        & $4^{1}$ & 7 & \circled{5} & 4 & 2\\ \cline{2-6}
        & 5 & \circled{5} & 4 & 2 & 0\\ \cline{2-6}
    \end{tabular}
    \subcaption{Iteration 4: Allocating the fourth unit.}
\end{minipage}
\caption{An illustration of progressive budgeting.} \label{tab:pb-illustration}
\end{table}

For the example we considered, the final seed set is $\{1^{1,2},3^1,4^1\}$ which is equivalent to $S_{1,2} \cup S_{3,1} \cup S_{4,1}$.

\section{Proof of Theorem~\ref*{thm:lower-bound}} \label{prf:thm:lower-bound}
We now prove Theorem~\ref*{lem:opt-substructure}.
\begin{proof}  The proof follows from the linearity of expectation in the definition of influence $\sigma$ and monotonicity.
\begin{align}
    \sigma(\cup_{i=1}^c S_i) &= \mathbb{E}\left[\sum_{v \in V} Y_T^{(v)} \bigg |  \bigcap_{v\in V} \{Y_0^{(v)} = \mathds{1}(v\in \cup_{i=1}^c S_i)\}  
    \right], \tag{from (\ref*{eq:equiv_def_influence})}\\
    &= \sum_{i=1}^c \mathbb{E}\left[\sum_{v \in V_i} Y_T^{(v)} \bigg |  \bigcap_{v\in V} \{Y_0^{(v)} = \mathds{1}(v\in \cup_{i=1}^c S_i)\} 
    \right], \tag{hard-partitioning; linearity of expectation}\\
    &\leq \sum_{i=1}^c \mathbb{E}\left[\sum_{v \in V_i} Y_T^{(v)} \bigg |  \bigcap_{v\in V} \{Y_0^{(v)} = \mathds{1}(v\in S_i)\} 
    \right], \tag{$\sigma_i$ is monotone non-decreasing}\\
    &=\sum_{i=1}^c \sigma_i(S_i). \nonumber 
\end{align}   
\end{proof}

\section{Proof of Lemma~\ref*{lem:opt-substructure}} \label{prf:lem:opt-substructure}
We now prove Lemma~\ref*{lem:opt-substructure}.
\begin{proof}
    The proof will follow by contradiction. Suppose there is some instance of Problem~\ref*{eq:ILP} and some cardinality $1\leq \ell < k$ such that starting with the budget allocation $\mathbf{k}^{*,(\ell+1)}$  for the optimal solution $S^{*,(\ell+1)}$ for budget $\ell+1$ and removing a unit of budget from any community results in a sub-optimal allocation (i.e. worse than $\mathbf{k}^{*,(\ell)}$).  For the special case that optimal solutions are unique for each cardinality and \texttt{sol-method} returns nested subsets, this condition simplifies to $S^{*,(\ell)} \not \subset S^{*,(\ell+1)}$.     
    
    If there is more than one optimal solution for budget $\ell$, fix any one as $S^{*,(\ell)}$.   Let us modify the budget allocation $\mathbf{k}^{*,(\ell+1)}$ of the optimal solution $S^{*,(\ell+1)}$ as follows.  Pick any community  $\tilde{i}\in\{1,\dots,c\}$ such that %
    \begin{align}
       k^{*,(\ell)}_{\tilde{i}} \leq k^{*,(\ell+1)}_{\tilde{i}}-1, \label{eq:prf:lem-sub:1}
    \end{align} that is even after removing a unit of budget for community  $\tilde{i}$ from allocation $\mathbf{k}^{*,(\ell+1)}$, there is still as much budget left over as there is for community  $\tilde{i}$  in the  allocation $\mathbf{k}^{*,(\ell)}$ (i.e. of the optimal solution $S^{*,(\ell)}$ for cardinality $\ell$).  Trivially since $\mathbf{k}^{*,(\ell+1)}$ allocates a larger budget overall, there must be one such community $\tilde{i}$ (if the solutions were nested, there would be exactly one).  Denote the corresponding modified solution  and budget allocation as  $\tilde{S}^{*,(\ell+1)}$ and $\tilde{\mathbf{k}}^{*,(\ell+1)}$ respectively.

    From our supposition, $\tilde{S}^{*,(\ell+1)}$ is not an optimal solution to Problem~\ref*{eq:ILP} for cardinality $\ell$ (its value is strictly worse than that of $S^{*,(\ell)}$).  We next consider constructing a solution of cardinality $\ell+1$ from $S^{*,(\ell)}$ by adding a unit of budget for community $\tilde{i}$.  With the nesting of the budget \eqref{eq:prf:lem-sub:1} for community $\tilde{i}$ specifically, by Assumption~\ref*{asmptn:decreasing} the resulting marginal gain must be at least as large as the marginal gain by adding a unit of budget to community $\tilde{i}$ for $\tilde{S}^{*,(\ell+1)}$.  (Recall from Remark~\ref*{remark:nested} that the assumption holds due to submodularity if the subsets chosen by \texttt{sol-method} are nested.)  Thus,  

    \begin{align}    &\hspace{-1cm}\overbrace{\left[\sum_{i=1}^c \sigma_i(S_{i,k^{*,(\ell)}_{i}}) \right] }^{\text{Value of } S^{*,(\ell)}}+ \overbrace{\left[  \sigma_{\tilde{i}}(S_{\tilde{i},k^{*,(\ell)}_{\tilde{i}}+1}) - \sigma_{\tilde{i}}(S_{\tilde{i},k^{*,(\ell)}_{\tilde{i}}}) \right]}^{\text{Marginal gain of augmenting } S^{*,(\ell)}} \nonumber \\
    &\geq \overbrace{\left[\sum_{i=1}^c \sigma_i(S_{i,k^{*,(\ell)}_{i}}) \right] }^{\text{Value of } S^{*,(\ell)}} + \overbrace{\left[  \sigma_{\tilde{i}}(S_{\tilde{i},k^{*,(\ell+1)}_{\tilde{i}}}) - \sigma_{\tilde{i}}(S_{\tilde{i},k^{*,(\ell+1)}_{\tilde{i}}-1}) \right]}^{\text{Marginal gain of augmenting } \tilde{S}^{*,(\ell+1)}}, \tag{by Assumption~\ref*{asmptn:decreasing} } \\
    &> \overbrace{\left[\sigma_{\tilde{i}}(S_{\tilde{i},k^{*,(\ell+1)}_{\tilde{i}}-1})  + \sum_{ \substack{i=1,\dots,c \\ i\neq \tilde{i}} } \sigma_i(S_{i,k^{*,(\ell+1)}_{i}}) \right]}^{\text{Value of } \tilde{S}^{*,(\ell+1)}} 
   + \overbrace{\left[  \sigma_{\tilde{i}}(S_{\tilde{i},k^{*,(\ell+1)}_{\tilde{i}}}) - \sigma_{\tilde{i}}(S_{\tilde{i},k^{*,(\ell+1)}_{\tilde{i}}-1}) \right]}^{\text{Marginal gain of augmenting } \tilde{S}^{*,(\ell+1)}}, \tag{$\tilde{S}^{*,(\ell+1)}$ is not optimal} \\
    &=\overbrace{\left[\sum_{i=1}^c \sigma_i(S_{i,k^{*,(\ell+1)}_{i}}) \right]}^{\text{Value of } S^{*,(\ell+1)}}. \nonumber
    \end{align}  Thus, the objective value of the optimal budget allocation $\mathbf{k}^{*,(\ell+1)}$ for a budget of $\ell+1$ is strictly less than the objective value of a budget allocation we constructed. This is a contradiction.  Thus our assumption about an instance lacking nesting (up to uniqueness) was incorrect.  
\end{proof}

\section{Computational complexity analysis} \label{sec:complexity} 
In this section, we analyze the computational complexity of the proposed framework (Algorithm~\ref*{alg:community-im}). The run-time of the proposed framework is the sum of the times taken at the three steps. It depends on the choice of community detection method as well as the solution method to solve IM for each community. We analyze the run-time involved at each step as follows.

\subsection{Learning the inherent community structure of the social network} 
The worst-case run-times of different community detection algorithms considered in this paper are given as follows: the Louvain method is $O(n \log n)$ \cite{blondel2008fast}, label propagation is $O(n+|E|)$ \cite{cordasco2010community}, and the Girvan-Newman method is $O(n|E|^2)$  \cite{zhu2002learning}.

\subsection{Generating candidate solutions by solving the influence maximization problem for each community} \label{sec:complexitystep2} 
If we use CELF++ to solve IM for $c$ different communities then we are solving $c$ problems of finding a $k$-node subset for each community from $n_i$ nodes, $i=1,\dots,c$. For the $i$th community, CELF++ iteratively builds the $k$-node subset as follows. First, find the best individual node by evaluating all $n_i$ subsets of cardinality one. Next, find the node with the highest marginal influence in the presence of the best individual node by evaluating (up to) all $n_i-1$ subsets of the previously selected best individual and an additional node. CELF++ then keeps adding nodes to the previous set in the same manner until the size of the current set is $k$. The number of $k$-node subsets evaluated at the $k$th step is $n_i-(k-1)$ in the worst case. Thus, 
the number of subsets evaluated in the worst case is 
\begin{align}
     & \hspace{-2cm} \text{ }\sum_{i=1}^{c}\left[n_i + (n_i-1) + \dots + (n_i-(k-1))\right] \nonumber\\
    &= nk - \frac{ck(k-1)}{2}. \label{eq:run-time-proposed}
\end{align}

On the contrary, if we use CELF++ for the entire network then the total number of subsets evaluated in the worst case is 
\begin{align}
n+(n-1)+\dots+(n-(k-1)) = nk - \frac{k(k-1)}{2}. \label{eq:run-time-celf++}
\end{align}

By comparing (\ref{eq:run-time-proposed}) and (\ref{eq:run-time-celf++}), we observe that the Generate-Candidates step of the proposed framework achieves a lower run-time compared to using the \texttt{sol-method} for the entire network by an additive factor of $(c-1)k(k-1)/2$. Furthermore, as $n_i \le n \ \forall i = 1,\dots, c$, the length of the diffusion while evaluating a subset of the nodes using Monte Carlo simulations within any community will always be smaller as compared to doing the same in the entire network. This further reduces the run-time of the Generate-Candidates step.

\subsection{Final seed set selection using progressive budgeting} 
The progressive budgeting method of final seed set selection solves `finding the maximum of $c$ elements' $k$ times. Hence, the worst-case run-time of progressive budgeting is $O(ck)$.

In practice, solving IM for each community (using a simulation-based \texttt{sol-method}) is the step that takes the most amount of time due to the costly Monte Carlo simulations. In that sense, the worst-case run-time of the proposed framework (with a simulation-based \texttt{sol-method}) to solve IM for each community is lower compared to the same for solving IM for the original network using the same simulation-based \texttt{sol-method}.
\end{document}